\documentclass[12pt]{article}%
\usepackage[nosort]{cite}
\usepackage{graphicx}
\usepackage{multicol}
\usepackage{amsfonts}
\usepackage{amssymb}
\usepackage{amsmath}
\usepackage{heck}
\usepackage{afterpage}
\usepackage{setspace}
\usepackage{verbatim}
\usepackage{color}
\usepackage{longtable}
\usepackage{float}
\usepackage{epsfig}
\usepackage{epstopdf}
\usepackage{adjustbox}
\usepackage{tikz}
\usepackage{tikz-cd}
\usepackage[margin=1in]{geometry}
\usepackage{titletoc}
\usepackage{hyperref}%
\usepackage{mathrsfs}
\usepackage{dsfont}
\usepackage{algorithm}
\usepackage{algpseudocode}
\usepackage{soul}
\usepackage{spectralsequences}
\usepackage{subcaption}

\providecommand{\U}[1]{\protect\rule{.1in}{.1in}}
\newsavebox{\mysavebox}

\hypersetup{colorlinks,citecolor=black,filecolor=black,linkcolor=black,urlcolor=black}
\usetikzlibrary{decorations.markings}

\numberwithin{equation}{section}

\newcommand{\be}{\begin{equation}}
\newcommand{\ee}{\end{equation}}

\usepackage{tikz}
\usetikzlibrary{arrows}
\usetikzlibrary{arrows.meta}
\usetikzlibrary{arrows,decorations.pathmorphing}
\usetikzlibrary{shapes.geometric,calc,arrows, positioning,shapes.misc,decorations.markings}
\tikzset{
  big arrow/.style={
    decoration={markings,mark=at position 1 with {\arrow[scale=2,#1]{>}}},
    postaction={decorate},
    shorten >=0.4pt},
  big arrow/.default=black}

\newcommand{\ba}{\begin{aligned}}
\newcommand{\ea}{\end{aligned}}

\usepackage{algorithmicx, algpseudocode,algorithm}
\usepackage{caption}

\usepackage{float}

\begin{document}

\date{July 2025}

\title{GSO Defects:\\[1mm] \Large{IIA/IIB Walls and the Surprisingly Stable $\mathrm{R}7$-Brane}}

\institution{PENN}{\centerline{$^{1}$Department of Physics and Astronomy, University of Pennsylvania, Philadelphia, PA 19104, USA}}
\institution{PENNmath}{\centerline{$^{2}$Department of Mathematics, University of Pennsylvania, Philadelphia, PA 19104, USA}}
\institution{CALTECH}{\centerline{$^{3}$Walter Burke Institute for Theoretical Physics, Caltech,
Pasadena, CA 91125, USA}}
\institution{IHES}{\centerline{$^{4}$Institut des Hautes Etudes Scientifiques, 91440 Bures-sur-Yvette, France}}
\institution{CERN}{\centerline{$^{5}$CERN, Theoretical Physics Department, 1211 Meyrin, Switzerland}}

\authors{
Jonathan J. Heckman,\worksat{\PENN,\PENNmath}\footnote{e-mail: \texttt{jheckman@sas.upenn.edu}}
Jacob McNamara,\worksat{\CALTECH}\footnote{e-mail: \texttt{jmcnamar@caltech.edu}}\\[4mm]
Julio Parra-Martinez,\worksat{\IHES}\footnote{e-mail: \texttt{julio@ihes.fr}} and
Ethan Torres\worksat{\CERN}\footnote{e-mail: \texttt{ethan.martin.torres@cern.ch}}
}

\abstract{The recently proposed Swampland Cobordism Conjecture predicts the existence of new non-supersymmetric objects which supplement
the spectrum of low-energy gravitational effective field theories. In this paper, we study a subset of these defects related to the GSO projection on the string worldsheet. These include the predicted domain wall between Type IIA and IIB superstring theories and the newly-discovered $\mathrm{R}7$-brane. We study these defects in two different ways: via long-string probes and target-space effective field theory. We find that the $\mathrm{R}7$-brane can be identified with a collapsed cylindrical configuration of the IIA/IIB wall, and further, that the $\mathrm{R}7$-brane is stable, in contrast to previous expectations. Moreover, we argue that BPS D-branes pulled across the IIA/IIB wall become non-BPS D-branes, which we identify with fluxbrane configurations. We show that the non-BPS D-branes of either Type II theory are charged under a $\mathbb{Z}_2$ remnant of the Ramond-Ramond potentials of the other, which we identify with the mod 2 reduction of the Ramond-Ramond fluxes. Similar considerations provide a complementary perspective on the Heterotic ${\mathfrak{so}(32)}$ S-duals of known non-BPS 7- and 8-branes in Type I string theory.}

{\small \texttt{\hfill CERN-TH-2025-136}}

\maketitle

\enlargethispage{\baselineskip}

\setcounter{tocdepth}{2}

\tableofcontents

\newpage

\section{Introduction} \label{sec:intro}

A basic but important issue which any putative theory of quantum gravity must
address is to identify the spectrum of states and objects whose existence affects physics at long distances. Indeed, this data is central in identifying the relevant degrees of freedom in any candidate UV completion. Celebrated objects of this sort are the BPS D-branes of open superstring theories \cite{Polchinski:1995mt}. As supersymmetric configurations, they are protected against quantum mechanical corrections, and thus robust statements can be made about their properties even at strong coupling.

While D-branes provide one example, there is no a priori guarantee that they provide a representative catalog of all possible extended objects in quantum gravity. Indeed, more general field configurations in gravitational effective field theories suggest a far broader range of possibilities, in line with the general expectation that UV-complete quantum gravity must include states of all possible charges \cite{Polchinski:2003bq}. In such situations, topological---rather than supersymmetric---considerations can provide a complementary perspective.

Recently, this approach was sharpened in the form of the Swampland Cobordism Conjecture \cite{McNamara:2019rup}.
This conjecture asserts that the bordism groups of quantum gravity are trivial:
\begin{equation}
\Omega_{k}^{\mathrm{QG}} = 0.
\end{equation}
Without a full definition of UV complete quantum gravity, we cannot directly check the validity of the Cobordism Conjecture. Nevertheless, this conjecture is still a powerful tool, since the bordism groups of spacetimes describable in a given low-energy effective field theory (EFT) are typically nontrivial. As such, we are able to \textit{predict} (sometimes \textit{postdict}) the existence of objects which enrich the spectrum of the original EFT and trivialize its bordism. For example, the Cobordism Conjecture correctly postdicts the existence of D-branes, O-planes, and other extended objects discovered earlier through other means. For some recent investigations of the Cobordism Conjecture, see references \cite{McNamara:2019rup, Montero:2020icj, Dierigl:2020lai, McNamara:2021cuo, Blumenhagen:2021nmi, Buratti:2021yia, Debray:2021vob, Andriot:2022mri, Dierigl:2022reg, Blumenhagen:2022bvh, Velazquez:2022eco, Angius:2022aeq, Blumenhagen:2022mqw, Angius:2022mgh, Blumenhagen:2023abk, Debray:2023yrs, Dierigl:2023jdp, Kaidi:2023tqo, Huertas:2023syg, Angius:2023uqk, Kaidi:2024cbx, Angius:2024pqk, Fukuda:2024pvu, Braeger:2025kra}.

More excitingly, the Cobordism Conjecture also predicts the existence of entirely new non-BPS objects! Notable examples include a codimension-one domain wall between the two Type II theories (the IIA/IIB wall) \cite{McNamara:2019rup}, codimension-two branes in Type IIB known as reflection 7-branes ($\mathrm{R}7$-branes) \cite{Dierigl:2022reg, Debray:2023yrs, Dierigl:2023jdp}, as well as a number of non-supersymmetric heterotic branes \cite{Kaidi:2023tqo, Kaidi:2024cbx, Fukuda:2024pvu}.\footnote{The $\mathrm{R}7$-brane, as well as the IIA/IIB wall, were predicted in passing in earlier work \cite{Distler:2009ri} for similar reasons.} In many of these examples, self-consistency of the candidate background requires these objects to have large string coupling near their core, so that they resist a microscopic treatment using worldsheet techniques (similarly to the supersymmetric $\mathrm{NS}5$-brane). Without simple worldsheet constructions or supersymmetric protection, these objects are harder to study than their better-known supersymmetric cousins.

In this paper, we aim to provide further evidence for the existence of some of these predicted objects and to establish some of their basic properties. The specific objects we consider are what we call \textit{GSO defects,} namely, defects in target space associated to the GSO projection on the string theory worldsheet. These include two basic sorts of object: The first are 8-brane \textit{GSO domain walls} (such as the IIA/IIB wall) which separate 10-dimensional string vacua that differ in the choice of GSO projection. The second are 7-brane \textit{GSO vortices} (such as the $F_L$ $\mathrm{R}7$-brane), characterized by a discrete holonomy in the target-space $\mathbb{Z}_2$ gauge field associated to the quantum symmetry of the GSO projection. Our main focus will be on the GSO defects in the Type II theories. Nevertheless, we will also consider GSO defects in the $\mathfrak{so}(32)$ Heterotic (Het$_{\mathfrak{so}}$) string theory, which turn out to be the S-duals of the well-known non-BPS $\mathrm{D}7$- and $\mathrm{D}8$-branes of type I string theory \cite{Sen:1998ki, Sen:1998rg, Sen:1998tt,  Witten:1998cd, Sen:1999mg}.

We study GSO defects from two complementary points of view. The first is in terms of the worldsheet EFT of the fundamental string. While GSO defects resist direct treatment in terms of the critical string worldsheet,\footnote{JM and JPM thank H. Ooguri, B. Rayhaun, S.-H. Shao, and A. Sharon for many previous discussions and collaboration on worldsheet constructions of vortices \cite{WorlsheetVortices}. JJH and ET thank M. Cveti\v{c} and G. Zoccarato for many attempts at realizing the IIA/IIB wall in supergravity, and M. Dierigl and M. Montero for attempts at realizing the $\mathrm{R}7$-brane directly in worldsheet terms \cite{Sappho}.} we may nevertheless consider the low-energy EFT describing small fluctuations of a long fundamental string in the presence of a GSO defect. Far away from the (potentially strongly-coupled) core of the defect, we can take the string coupling to be small, so that the long-string EFT is under control. In this regime, we are able to characterize some aspects of the IIA/IIB wall and the $\mathrm{R}7$-brane in terms of defects in the long-string EFT.

Our second perspective is to consider the behavior of GSO defects in the target space and their effect on fields and probe branes. In particular, we find that BPS $\mathrm{D}p$-branes on the IIA (resp. IIB) side of the wall are related to non-BPS $\mathrm{D}p$-branes on the IIB (resp. IIA) side of the wall. As we show, these non-BPS $\mathrm{D}p$-branes turn out to be charged under a $\mathbb{Z}_2$ remnant of the odd-form (resp. even-form) RR potentials present for their BPS counterparts. We further argue that the non-BPS D$p$-branes are in turn just the initial configuration for the non-supersymmetric flux-$p$-branes of references \cite{Gutperle:2001mb, Emparan:2001gm, Cvetic:2023plv}. Note that this result contradicts previous statements in the literature (e.g., \cite[Section 3.3]{Schwarz:1999vu}) that the non-BPS D$p$-branes carry no conserved charge and can decay into purely neutral radiation.\footnote{There is no contradiction between our claim and the $K$-theory classification of D-branes; indeed, our arguments in Appendices \ref{app:MOD2} and \ref{app:FLUXBRANE} for this $\mathbb{Z}_2$ charge rely on $K$-theory. Our claim is that the non-BPS D-branes carry charge associated to the torsional holonomies of the differential $K$-theory field, rather than the integrally quantized fluxes, as is the case for the BPS D-branes.}

\begin{figure}[t!]
    \centering
\includegraphics[trim={0 2cm 0 0cm},clip,width=12cm,scale=1]{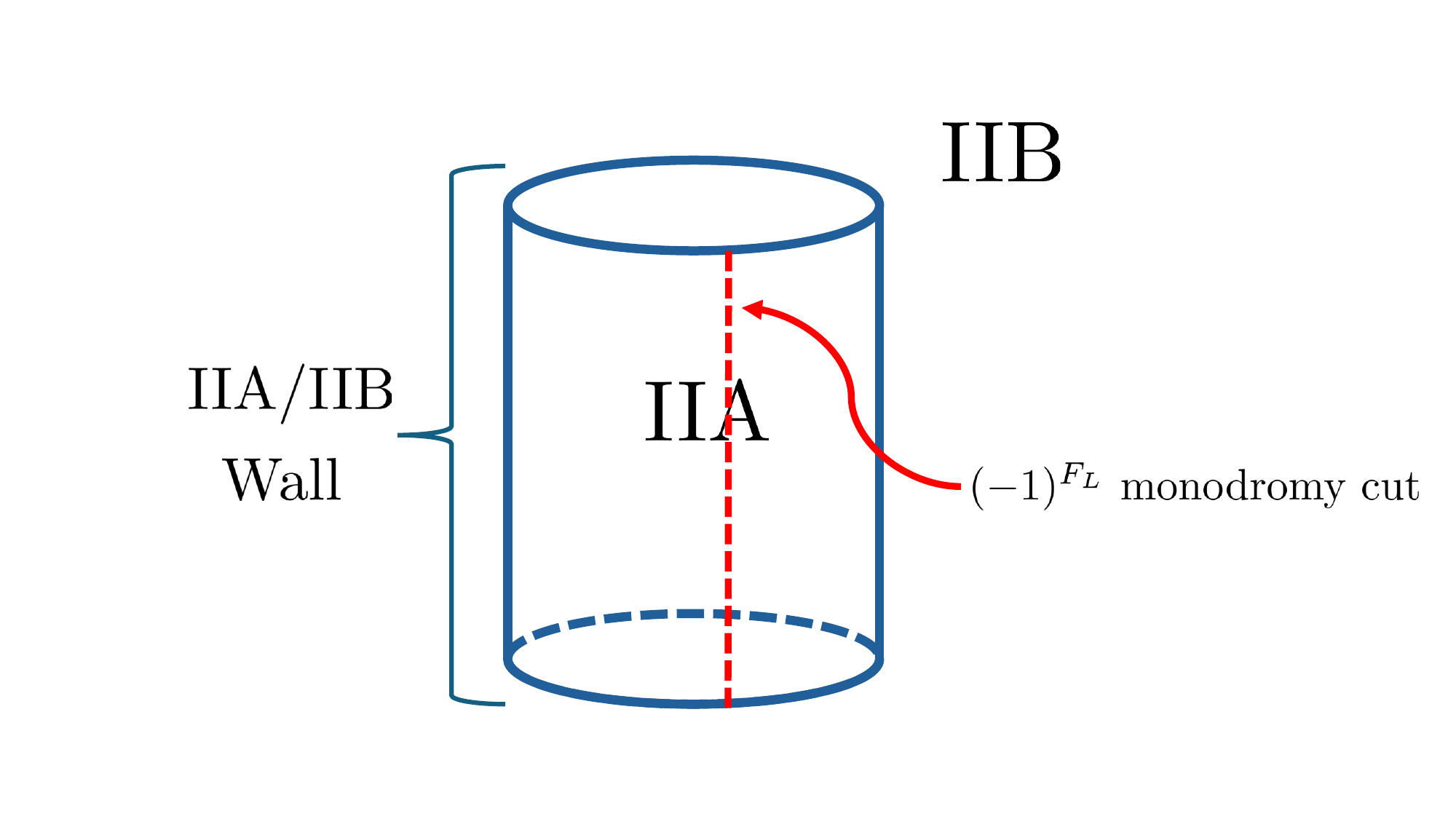}
    \caption{Configuration of a IIA/IIB wall with spatial topology $S^1\times \mathbb{R}^{7}$ and a $(-1)^{F_L}$ monodromy around its 1-cycle, with IIA on the inside and IIB on the outside.}
    \label{fig:ABwallIntro}
\end{figure}

One aspect of the IIA/IIB wall we are able to infer is that it serves as a Neumann boundary condition for the $\mathbb{Z}_2$ gauge field for left-moving spacetime fermion number $(-1)^{F_L}$ (on both sides, so that the gauge fields fluctuate independently). As a result, we may consider a configuration in which the IIA/IIB wall is wrapped into a cylinder extended in 7 spatial directions with a discrete holonomy of $(-1)^{F_L}$ around a circle in the two transverse directions (see Figure \ref{fig:ABwallIntro}). Due to the tension of the IIA/IIB wall, this configuration will be unstable, and the cylinder will dynamically shrink. Viewed from far away, this configuration is indistinguishable from the $F_L$ $\mathrm{R}7$-brane of the Type II theory on the outside, and so we conjecture that the endpoint of this shrinking process is simply the $\mathrm{R}7$-brane itself.

We give two further cross-checks of this picture. First, we are able to match the worldvolume form fields on the $F_L$ $\mathrm{R}7$-brane in either Type II theory with the bulk Ramond-Ramond (RR) form fields of the other. This suggests that these form fields on the $\mathrm{R}7$-brane arise as the zero modes of the RR fields of the other Type II theory living inside a small cylindrical IIA/IIB wall. Second, we verify that an analogous transition between a GSO domain wall and a GSO vortex is possible in the Het$_{\mathfrak{so}}$ theory, at least at strong coupling where we may use known D-brane physics in the dual Type I description.

Perhaps counterintuitively, our considerations suggest that the $\mathrm{R}7$-brane is actually \textit{stable}. 
As an object of low codimension, it suffers a potential instability to spreading out in the transverse directions into a configuration of lower energy density (see e.g., \cite{Adams:2001sv}). However, if this were possible, the resulting configuration should be describable in terms of a smooth field configuration in the low-energy EFT. As there is no smooth configuration of the Type II supergravity EFT realizing the charge of the $\mathrm{R}7$-brane, we conclude that the $\mathrm{R}7$-brane must indeed be stable. This criterion, that an object whose charge cannot be realized by a smooth EFT configuration must be stable, was emphasized in \cite{Kaidi:2024cbx}.\footnote{In \cite{Kaidi:2024cbx}, the authors introduced the SUGRA bordism groups $\Omega_k^{\rm SUGRA}$ to characterize the charges that cannot be realized by smooth EFT configurations. Indeed, the $\mathrm{R}7$-branes are associated to nontrivial classes in the SUGRA bordism groups of the respective Type II theories.}

To build further intuition for the stability of the $\mathrm{R}7$-brane, it is helpful to contrast it with the GSO vortex of the Het$_{\mathfrak{so}}$ theory. As reviewed below, this GSO vortex is given by the S-dual of the Type I non-BPS 7-brane first identified in \cite{Witten:1998cd}, which is associated to a discrete $KO$-theory charge. This 7-brane has a worldvolume tachyon, and suffers a spreading-out instability \cite{Frau:1999nyc, Loaiza-Brito:2001yer}. The end result of this instability is a smooth configuration of the ${\rm Spin}(32)/\mathbb{Z}_2$ gauge field with nonzero magnetic flux. This smooth configuration is able to realize the holonomy of the non-BPS 7-brane because the associated $\mathbb{Z}_2$ gauge field is embedded in the center of ${\rm Spin}(32)/\mathbb{Z}_2$. In contrast, the $\mathbb{Z}_2$ gauge field $(-1)^{F_L}$ in either Type II theory is not embedded inside any continuous (internal) gauge group, and thus the $\mathrm{R}7$-brane cannot dissolve into a dilute gauge flux. This is analogous to the deep IR of a superconducting phase where a $U(1)$ gauge field is broken to $\mathbb{Z}_2$, and the corresponding flux tubes are stable string excitations.

The rest of this paper is organized as follows. First, in Section \ref{sec:BACKGROUND}, we review the relevant features of the GSO projection, as well as our two main examples of Type II and Het$_{\mathfrak{so}}$ string theory. In Section \ref{sec:LONGSTRING}, we study GSO walls from the perspective of a long-string, and in Section \ref{sec:TARGET}, we give a target space analysis. In Section \ref{sec:VORTICES} we turn to codimension-two GSO vortices, and discuss their relationship to cylindrical configurations of GSO walls. Section \ref{sec:STABILITY} discusses the stability of the different GSO defects. Finally, Section \ref{sec:CONC} contains our conclusions and some directions for future investigation.

We supplement our main discussion with a number of appendices, containing important calculations and arguments to support our main points. In Appendix \ref{app:BACKFIRING}, we detail subtleties in the GSO projection arising from discrete anomalies (or ``backfiring bosonization'') \cite{BoyleSmith:2024qgx}. Appendix \ref{app:EXTENSION} discusses the origin of the extension of the Type I gauge group. Appendix \ref{app:MOD2} derives the relationship between the $\mathbb{Z}_2$ remnants of the RR potentials of type IIA (resp. IIB) and the mod $2$ reductions of the RR fluxes in type IIB (resp. IIA). Appendix \ref{app:LAGMULT} reviews the path integral formulation of RR fluxes in terms of integrals over flux configurations with suitable Lagrange multipliers. In Appendix \ref{app:FLUXBRANE} we present some general comments on fluxbranes, and argue that non-BPS D-branes can be precisely identified with RR fluxbranes. Finally, Appendix \ref{app:TYPEI} provides some additional details on the $\mathrm{D}7$-brane of type I string theory, and argues for the connection between the D7-brane and a cylindrical D8-brane.

\section{Review of the GSO Projection}\label{sec:BACKGROUND}

We begin by reviewing the basic structure of the GSO projection for the Ramond-Neveu-Schwarz (RNS) string. In the RNS formalism, the physical string spectrum is obtained only after performing a projection onto states of definite fermion number in the Neveu-Schwarz (NS) and Ramond (R) sectors. Among other reasons, this projection is required in order to eliminate the tachyonic ground state in the NS sector and have a consistent string perturbation theory. Roughly speaking, the GSO projection arises from the sum over possible spin structures $\sigma$ on the superstring worldsheet, or, essentially equivalently, from gauging worldsheet fermion parity $(-1)^f$. There are a number of technical subtleties in the precise meaning of the sum over spin structures (see e.g., \cite{BoyleSmith:2024qgx,Witten:2012bh}) which will not play any role in our main discussion, but which we comment on in Appendix \ref{app:BACKFIRING}.

In performing a sum over worldsheet spin structures, one is free to include a phase factor $(-1)^{{\rm Arf}(\sigma)}$, where ${\rm Arf}(\sigma)$ is the Arf invariant of the spin structure $\sigma$. Including this factor corresponds to stacking a Majorana chain \cite{Kitaev:2000nmw} onto the worldsheet \cite{Kaidi:2019pzj,Kaidi:2019tyf,Witten:2023snr}. The impact of this factor is to shift the fermion number of the Ramond sector ground state, or equivalently, to modify the GSO projection to project onto states of opposite fermion number in the Ramond sector. One may view the choice of including or not including the factor $(-1)^{{\rm Arf}(\sigma)}$ as a discrete theta angle of the worldsheet theory, which should correspond in target space to a discrete parameter labeling string vacua that differ only in the choice of GSO projection.

After performing the GSO projection, the resulting theory admits a $\mathbb{Z}_2$ quantum symmetry which assigns charge to states in the Ramond sector. This $\mathbb{Z}_2$ global symmetry of the worldsheet theory corresponds to a $\mathbb{Z}_2$ gauge symmetry in target space. As reviewed below, the details of this gauge symmetry depend on the particular theory considered; for the sake of exposition, we will refer to this quantum symmetry here as $(-1)^{F_{\rm GSO}}$. An essential technical fact is that the operation of stacking $(-1)^{{\rm Arf}(\sigma)}$ before taking the GSO projection corresponds, after GSO, to a further gauging of the quantum symmetry $(-1)^{F_{\rm GSO}}$ (See e.g., \cite{Shao:2023gho}).\footnote{When bosonization ``backfires'' (see Appendix \ref{app:BACKFIRING}), this fact remains true provided that the further gauging of $(-1)^{F_{\rm GSO}}$ is done with the opposite refermionization procedure as was used in the GSO projection. See \cite[Section 4.3]{BoyleSmith:2024qgx} for details.} This can be understood as follows: gauging $(-1)^{F_{\rm GSO}}$ projects out the physical string states in the Ramond sector, and adds in new twisted-sector states which correspond to the Ramond sector states of opposite worldsheet fermion parity.

Another key effect has to do with the possibility of a mixed anomaly between worldsheet fermion number $(-1)^f$ and some continuous global symmetry $G$ of the worldsheet theory. The specific mixed anomaly is characterized by a 3d action
\begin{equation}\label{eq:extension_anomaly}
    \int_{M_3} a\cup \rho^* e(g),
\end{equation}
where $\Sigma_2 = \partial M_3$ is the worldsheet, $a\in H^1(M_3,\mathbb{Z}_2)$ is the $(-1)^f$ background field, $e(g)\in H^2(G, \mathbb{Z}_2)$, and $\rho:M_3\rightarrow BG$ is the background field for $G$\cite{Tachikawa:2017gyf}. In the presence of such an anomaly, the global symmetry after GSO is given by a central extension $\hat{G}$ of $G$ by the quantum symmetry $(-1)^{F_{\rm GSO}}$ of GSO,
\begin{equation}
    \mathbb{Z}_2^{F_{\rm GSO}} \to \hat{G} \to G.
\end{equation}
When this happens, the target space gauge symmetry corresponding to $(-1)^{F_{\rm GSO}}$ is embedded as a central $\mathbb{Z}_2$ subgroup of the continuous $\hat{G}$ gauge symmetry.

\subsection{Example: Type II String Theory}\label{sec:Type_II_review}

In this paper, we consider defects related to the GSO projection in two main examples. The first example is Type II string theory. In the lightcone-gauge RNS formalism, Type II string theory is described by worldsheet fields $(X^i, \psi^i, \widetilde{\psi}^i)$, where $i=1,...,8$ parametrize eight transverse spatial directions, $X^i$ are 2d bosons, and $\psi^i$ and $\widetilde{\psi}^i$ are 2d left- and right-moving Majorana-Weyl fermions respectively. The action on a worldsheet $\Sigma$ is simply
\begin{equation}
    S = \int_{\Sigma}\left(\partial X^i\overline{\partial}X^i+\psi^i\overline{\partial}\psi^i+\widetilde{\psi}^i\partial \widetilde{\psi}^i\right).
\end{equation}
In Type II string theory, one must sum over independent spin structures $(\sigma_L, \sigma_R)$ for the left- and right-moving fermions respectively, and one may choose to include a factor of the Arf invariant for either chirality. Our conventions will be that Type IIB string theory is obtained by not including this factor for either spin structure, while Type IIA string theory includes this factor only for the left-moving spin structure $\sigma_L$.\footnote{The two other possibilities (including this factor on both sides or only on the right-moving side) yield theories equivalent to Type IIB and Type IIA upon performing a spacetime reflection.}

After performing the GSO projection, there are two independent quantum symmetries $(-1)^{F_L}$ and $(-1)^{F_R}$ which assign charge to the left- and right-moving Ramond sectors respectively. Due to a mixed anomaly of the form \eqref{eq:extension_anomaly} between worldsheet fermion number and rotations in the eight transverse directions, the diagonal symmetry $(-1)^{F} = (-1)^{F_L} (-1)^{F_R}$ is embedded as the center of the transverse rotation group ${\rm Spin}(8)$ and corresponds to target-space fermion number. Thus, the only discrete $\mathbb{Z}_2$ gauge symmetry in the Type II target space arising from GSO corresponds to either of $(-1)^{F_L}$ or $(-1)^{F_R}$, which by convention we will take to be $(-1)^{F_L}$. The discrete parameter distinguishing Type IIA from Type IIB and the $\mathbb{Z}_2$ gauge field $(-1)^{F_L}$ are the $H^0$ and $H^1$ components of the twisting of $K$-theory considered in \cite{Distler:2009ri}.

\subsection{Example: Het$_\mathfrak{so}$ String Theory}\label{sec:het_review}

The second example we consider is the Het$_{\mathfrak{so}}$ heterotic string theory. In the lightcone-gauge RNS formalism, the fermionic construction of the heterotic string involves worldsheet fields $(X^i, \widetilde{\psi}^i, \lambda^A)$, where $X^i, \widetilde{\psi}^i$ are as in the Type II string, and $\lambda^A$ for $A = 1, \dots, 32$ are 32 left-moving worldsheet fermions. The action is
\begin{equation}
    S = \int_{\Sigma}\left(\partial X^i\overline{\partial}X^i+\lambda^A\overline{\partial}\lambda^A+\widetilde{\psi}^i\partial \widetilde{\psi}^i\right).
\end{equation}
In order to obtain the Het$_\mathfrak{so}$ heterotic string, one must again sum over left- and right-moving spin structures $(\sigma_L, \sigma_R)$ independently. The right-handed GSO projection behaves similarly to the Type II string, and its quantum symmetry corresponds to target-space fermion number $(-1)^F$ in the ten non-compact spacetime dimensions. Our focus will be on the left-handed GSO projection which acts on the 32 left-moving fermions $\lambda^A$. The result of the left-handed GSO projection is a Narain CFT obtained by the compactification of 16 chiral bosons, $\varphi_i$, on a 16-dimensional even self-dual lattice $\Gamma_{16}$. 

In order to describe the effect of stacking the factor $(-1)^{{\rm Arf}(\sigma_L)}$, it will be useful to consider the global rotation symmetry acting on the 32 fermions. Before GSO, we have an ${\rm SO}(32)$ global symmetry under which the 32 fermions $\lambda^A$ transform as a vector. This symmetry has a mixed anomaly with left-handed fermion number $(-1)^{f_L}$ of the form \eqref{eq:extension_anomaly}, and thus will be extended by the quantum symmetry of the left-handed GSO projection. Moreover, left-handed worldsheet fermion number $(-1)^{f_L}$ can be identified with the central element of ${\rm SO}(32)$. As all states with nonzero $(-1)^{f_L}$ charge are projected out by GSO, the worldsheet global symmetry after GSO will involve a quotient as well.

Putting these two effects together, we learn that the worldsheet global symmetry after the left-handed GSO projection is given by ${\rm Spin}(32)/\mathbb{Z}_2$, where $\mathbb{Z}_2$ corresponds to a central element in ${\rm Spin}(32)$ that lifts the unique central element of ${\rm SO}(32)$. As it turns out, there are two such elements in ${\rm Spin}(32)$. The center of ${\rm Spin}(32)$ is given by $\mathbb{Z}_2^s \times \mathbb{Z}_2^c$, where the generators $(-1)^s$ and $(-1)^c$ of $\mathbb{Z}_2^s$ and $\mathbb{Z}_2^c$ act nontrivially in the spinor and conjugate-spinor representations respectively. Both elements $(-1)^s$ and $(-1)^c$ lift the central element of ${\rm SO}(32)$, and so there are two nominal possibilities for the global symmetry of the heterotic string worldsheet after GSO, given by ${\rm Spin}(32)/\mathbb{Z}_2^s$ and ${\rm Spin}(32)/\mathbb{Z}_2^c$. These two groups are abstractly isomorphic, and the two options are exchanged by a reflection in one of the bosonized fields $\varphi_i \to - \varphi_i$.

Nevertheless, if we fix the action of ${\rm Spin}(32)$ on the Ramond sector before GSO projection, it is meaningful to discuss the difference between the options ${\rm Spin}(32)/\mathbb{Z}_2^s$ and ${\rm Spin}(32)/\mathbb{Z}_2^c$. The states which transform as spinors under this internal rotation group are given by massive excitations of the heterotic string, arising from quantization of the zero modes of $\lambda^A$ in the Ramond sector. In order to specify the action of ${\rm Spin}(32)$ on the Ramond sector, one must choose which of $(-1)^s$ and $(-1)^c$ acts as left-moving fermion parity $(-1)^{f_L}$; the two options are exchanged upon stacking the factor $(-1)^{{\rm Arf}(\sigma_L)}$. Depending on this choice, the states which survive the GSO projection live in the weight lattice of either ${\rm Spin}(32)/\mathbb{Z}_2^s$ or ${\rm Spin}(32)/\mathbb{Z}_2^c$, both of which are abstractly isomorphic to $\Gamma_{16}$. Our convention will be that without the factor $(-1)^{{\rm Arf}(\sigma_L)}$ one obtains the global form ${\rm Spin}(32)/\mathbb{Z}_2^c$, which allows for the spinor representation, while with the factor $(-1)^{{\rm Arf}(\sigma_L)}$ one obtains ${\rm Spin}(32)/\mathbb{Z}_2^s$, allowing for the conjugate-spinor representation.

\subsubsection{The S-Dual Type I Theory}\label{sec:Type_I_review}

Finally, it is also of interest to consider the avatar of the Het$_\mathfrak{so}$ GSO projection in the S-dual Type I string theory. Type I string theory is given by an orientifold of the Type IIB string by worldsheet parity $\Omega$ together with the introduction of 32 space-filling D9-branes in order to cancel the charge of the O9-plane. In order to match with the above discussion, we must understand why the gauge group of the Type I string theory is given by $\mathrm{Spin}(32)/\mathbb{Z}_2$, and what determines the chirality of the $\mathbb{Z}_2$ subgroup in the quotient.

At first glance, the gauge group of Type I appears to be $\mathrm{O(32)}$, as this is the $\Omega$-invariant subgroup of the $\mathrm{U}(32)$ gauge group living on the stack of 32 D9-branes in Type IIB. Let us first consider the identity component $\mathrm{SO}(32)$. The orientifold projection removes the continuous gauge potential $B_2$ from the spectrum, but leaves behind a discrete $\mathbb{Z}_2$ 2-form gauge field under which the fundamental string is charged. In Type II string theory, in the presence of D-branes, the D-brane localized gauge fields $F_{\mathrm{D}p}$ transform as
\begin{equation}
    F_{\mathrm{D}9} \to F_{\mathrm{D}9} - \frac{1}{2 \pi} d \Lambda
\end{equation}
under a bulk gauge transformation $B_2 \to B_2 + d \Lambda$, where we use the embedding ${\rm U}(1) \hookrightarrow {\rm U}(N)$ for a stack of $N$ D-branes. As a result, in Type I string theory, the discrete remnant of the $B_2$ field gauges the $\mathbb{Z}_2$ center symmetry of the $\mathrm{SO}(32)$ gauge field, reducing the physical gauge group to the quotient $\mathrm{SO}(32)/\mathbb{Z}_2$. 

That being said, the whole gauge group is known to lift to $\mathrm{Spin}(32) / \mathbb{Z}_2$. Let us briefly discuss how this comes about. In Type I string theory there is also a $\mathbb{Z}_2$ 8-form gauge potential given by the discrete remnant of the RR field $C_8$, which transforms as $C_8 \to -C_8$ under $\Omega$. Now, in Type IIB, before the orientifold projection, we have a coupling of the form:
\begin{equation}\label{eq:C8_coupling}
    \int C_8 \wedge {\rm Tr}(F_{\mathrm{D}9}),
\end{equation}
on the D9-brane worldvolume, between $C_8$ and the $\mathrm{U}(32)$ gauge fields. Globally, this coupling is better described as a coupling between $C_8$ and the characteristic class $c_1$ of the $\mathrm{U}(32)$ gauge bundle. In Type I, the coupling \eqref{eq:C8_coupling} descends to a coupling between the $\mathbb{Z}_2$ remnant of $C_8$ and the characteristic class $w_2$ of the $\mathrm{SO}(32)$ gauge bundle, using the fact that $c_1$ pulls back to $w_2$ under the embedding $\mathrm{SO}(32) \hookrightarrow \mathrm{U}(32)$. As a result, the remnant of $C_8$ gauges the magnetic 7-form symmetry of $\mathrm{SO}(32)$ gauge theory, extending $\mathrm{SO}(32)$ to $\mathrm{Spin}(32)$. The new 1-form $\mathbb{Z}_2$ gauge field that extends $\mathrm{SO}(32)$ is given by the electromagnetic dual of the discrete remnant of $C_8$. The identity of this dual field is surprisingly subtle, and is described in Appendix \ref{app:EXTENSION}.

Taking both effects into account, we conclude that the global form of the Type I gauge group is given by $\mathrm{Spin}(32)/\mathbb{Z}_2$. In order to address the chirality of the $\mathbb{Z}_2$ quotient in $\mathrm{Spin}(32)/\mathbb{Z}_2$, let us return to the disconnected component of $\mathrm{O}(32)$, consisting of internal reflections. As explained in \cite{Witten:1998cd}, the disconnected component of $\mathrm{O}(32)$ is broken by D$(-1)$-brane instantons, whose partition function picks up a sign under an internal reflection. Equivalently, the disconnected component is broken by the vev of $C_0$, which controls the phase of the D$(-1)$-brane partition function. There are two values of $C_0$ invariant under the orientifold projection $C_0 \to -C_0$, given by $C_0 = 0$ or $C_0 = \pi$, taking into account the identification $C_0 \sim C_0 + 2\pi$. As explained in \cite{BergmanSeminar, Montero:2022vva}, the two choices are gauge equivalent, as they are exchanged under an internal reflection in $\mathrm{O}(32)$. Thus, the chirality of the $\mathbb{Z}_2$ quotient, though ultimately unphysical, is controlled by the nominal vev of $C_0$.

We now track the effect of this parameter on the worldsheet of the Type I $\mathrm{D}1$-string, the S-dual of the Het$_{\mathfrak{so}}$ fundamental string. Open strings in the $1-9$ sector transform in the fundamental representation of $\mathrm{SO}(32)$. Following the discussion in \cite{Polchinski:1995df} (see also \cite{Polchinski:1998rr}), we consider a wrapped D-string on a long circle with periodic boundary conditions for the fermionic $1-9$ strings. Observe that the zero modes for these fermionic modes $\Lambda^{i}_{0}$ (with $i = 1,...,32$)  satisfy the Clifford algebra $\{\Lambda_{0}^{i}, \Lambda_{0}^{j} \} = 2 \delta^{ij}$, and so the corresponding ground states on the $\mathrm{D}1$-string build up spinor representations of $\mathrm{Spin}(32) / \mathbb{Z}_2$, just as in the Het$_{\mathfrak{so}}$ string. Turning on a vev $C_0 = \pi$ shifts the fundamental IIB string charge of the $\mathrm{D}1$-string, altering the projection induced by the $\mathbb{Z}_2$ remnant of the $B_2$ field. As \cite{Montero:2022vva} shows, the effect is to introduce a discrete theta angle in the $\mathrm{D}1$-string worlsheet gauge theory, which changes the chirality of the spinors of the ground state associated with the wrapped $\mathrm{D}1$-string. Turning on this discrete theta angle on the $\mathrm{D}1$-string is the S-dual of stacking a factor $(-1)^{{\rm Arf}(\sigma_L)}$ on the Het$_{\mathfrak{so}}$ string worldsheet.

\section{GSO Domain Walls: Long String Probe} \label{sec:LONGSTRING}

Motivated by the Swampland Cobordism conjecture, our aim in this section is to characterize domain walls connecting pairs of perturbative string vacua differing only by the choice of whether or not to include a factor of $(-1)^{{\rm Arf}(\sigma)}$ in the GSO projection. We focus on cases where far from the wall we have asymptotically supersymmetric Minkowski vacua. Moreover, we will set the asymptotic value of the string coupling $g_s$ to some small fixed value far away from the domain wall on each side. As mentioned above, our two main examples live in Type II and Het$_{\mathfrak{so}}$ string theory. These are given by the domain wall that separates Type IIA from Type IIB string theory and the domain wall between the two equivalent versions of the Het$_\mathfrak{so}$ theory with nominal gauge groups ${\rm Spin}(32)/\mathbb{Z}_2^s$ and ${\rm Spin}(32)/\mathbb{Z}_2^c$.

\begin{figure}[t!]
\centering
\includegraphics[scale=0.4, trim={0 2cm 0 2cm}]{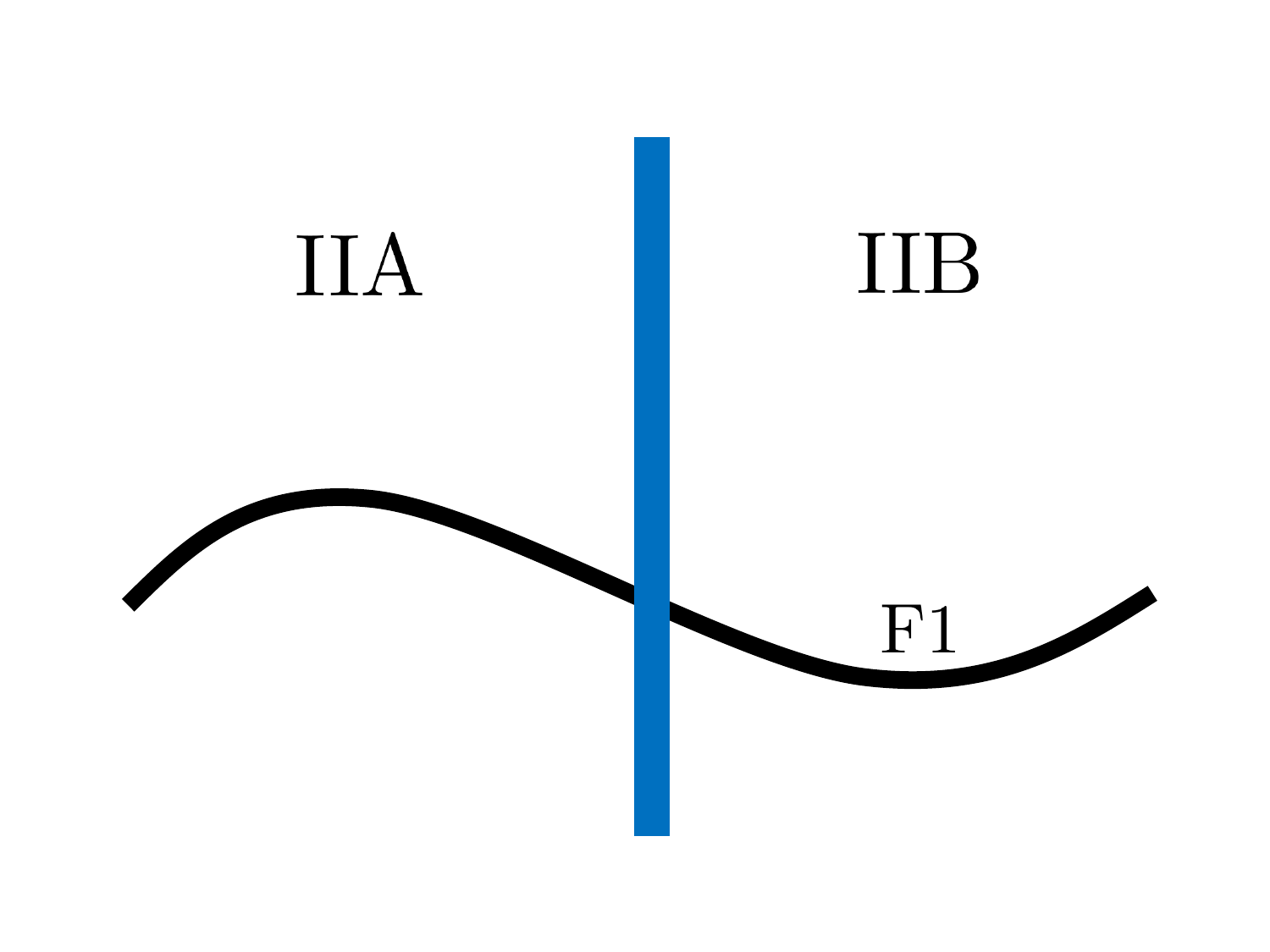}
\caption{Target space picture of a long $\mathrm{F}1$-string crossing the IIA/IIB wall.}
\label{fig:f1crossing}
\end{figure}

\begin{figure}[t!]
    \centering
\includegraphics[trim={0 0cm 0 0cm},clip,width=12cm,scale=0.8]{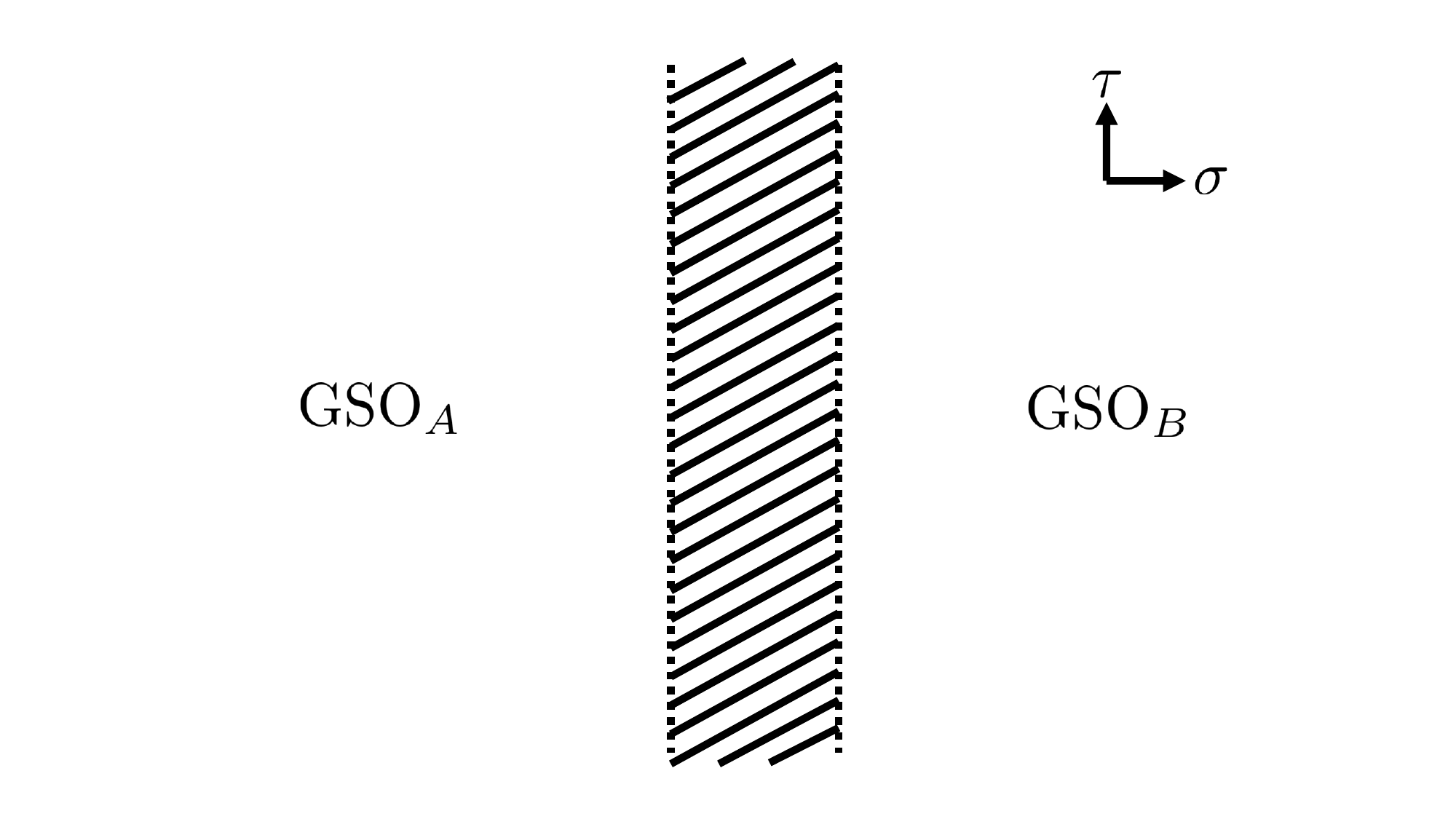}
    \caption{Depiction of a GSO interface in type II theory for a long string. In the long string, $g_s \rightarrow 0$ limit we can
    treat this as a fixed defect in which the choice of GSO projection is different on the two sides of the wall. We have depicted this as a
    ``shaded region'' to emphasize that we are remaining agnostic as to the microscopic structure of this interface. That being said, anomaly inflow considerations indicate that anomalies can be cancelled by a 1D Majorana fermion trapped on a thin interface. }
    \label{fig:InterfaceGSO}
\end{figure}

In general, we do not expect to have a worldsheet construction of GSO domain walls in perturbative string theory, as the dilaton may have a strong gradient in the vicinity of the wall.\footnote{Compare with the supersymmetric $\mathrm{NS}5$-brane, whose core does not admit a simple worldsheet description. This issue also arises for the non-supersymmetric heterotic branes considered in \cite{Kaidi:2023tqo, Kaidi:2024cbx, Fukuda:2024pvu}.} Rather than produce a full worldsheet construction, we shall instead study these domain walls under one crucial assumption: that a single long fundamental string may be stretched across the domain wall, as depicted in Figure \ref{fig:f1crossing}. This assumption is not justified from first principles. However, we will see that it leads to a self-consistent picture for the physics of the domain wall. While it is difficult to study the interaction of critical strings with the domain wall, in the regime of a single long string we may instead consider the low-energy effective description of small excitations of the long string. At scales longer than the thickness of the domain wall, its imprint must be describable by some interface in the long-string effective theory (see  Figure \ref{fig:InterfaceGSO}). By characterizing some universal features of this interface, we will be able to predict features of the GSO domain walls.

As it turns out, there is a well-known topological interface between worldsheet theories that differ by the factor $(-1)^{{\rm Arf}(\sigma_L)}$. Before the GSO projection, this interface can be described in terms of a single unpaired Majorana that lives at the end of the Majorana chain \cite{Kitaev:2000nmw}. This is a gapped interface, as the (anomalous) partition function of a single 1D Majorana mode is independent of the metric, and is given by either $\sqrt{2}$ or $0$ depending on whether the boundary conditions are antiperiodic or periodic \cite{Witten:2023snr, WittenTalkOne, WittenTalkTwo, Freed:2024apc}. Alternatively, after GSO, this interface can be described as the interface obtained by gauging the quantum symmetry of GSO on half of the worldsheet. As described above, this corresponds in Type II to the gauging interface for $(-1)^{F_L}$, and in $\mathrm{Het}_{\mathfrak{so}}$ to the gauging interface for the $\mathbb{Z}_2$ center of either form of ${\rm Spin}(32)/\mathbb{Z}_2$.

It seems highly unlikely that the interface in the long-string EFT corresponding to the GSO domain wall is simply given by this topological interface. Instead, our expectation is that the GSO domain wall corresponds to some non-universal, non-topological deformation of this topological interface. Nevertheless, we expect that certain robust features controlled by topology such as anomaly inflow and the spectrum of boundary states and (quasi-)stable objects can be reliably extracted from these coarse considerations.

In particular, one defining property of the topological gauging interface is that the topological line operators implementing the quantum symmetry of GSO may end on the gauging interface.
The quantum symmetry will be a global symmetry in the ungauged region. The corresponding topological line operator can end on the gauging interface simply because it becomes indistinguishable from the identity operator in the gauged region. 

\begin{figure}[t!]
    \centering
\includegraphics[trim={0 2cm 0 2cm},clip,width=12cm,scale=0.8]{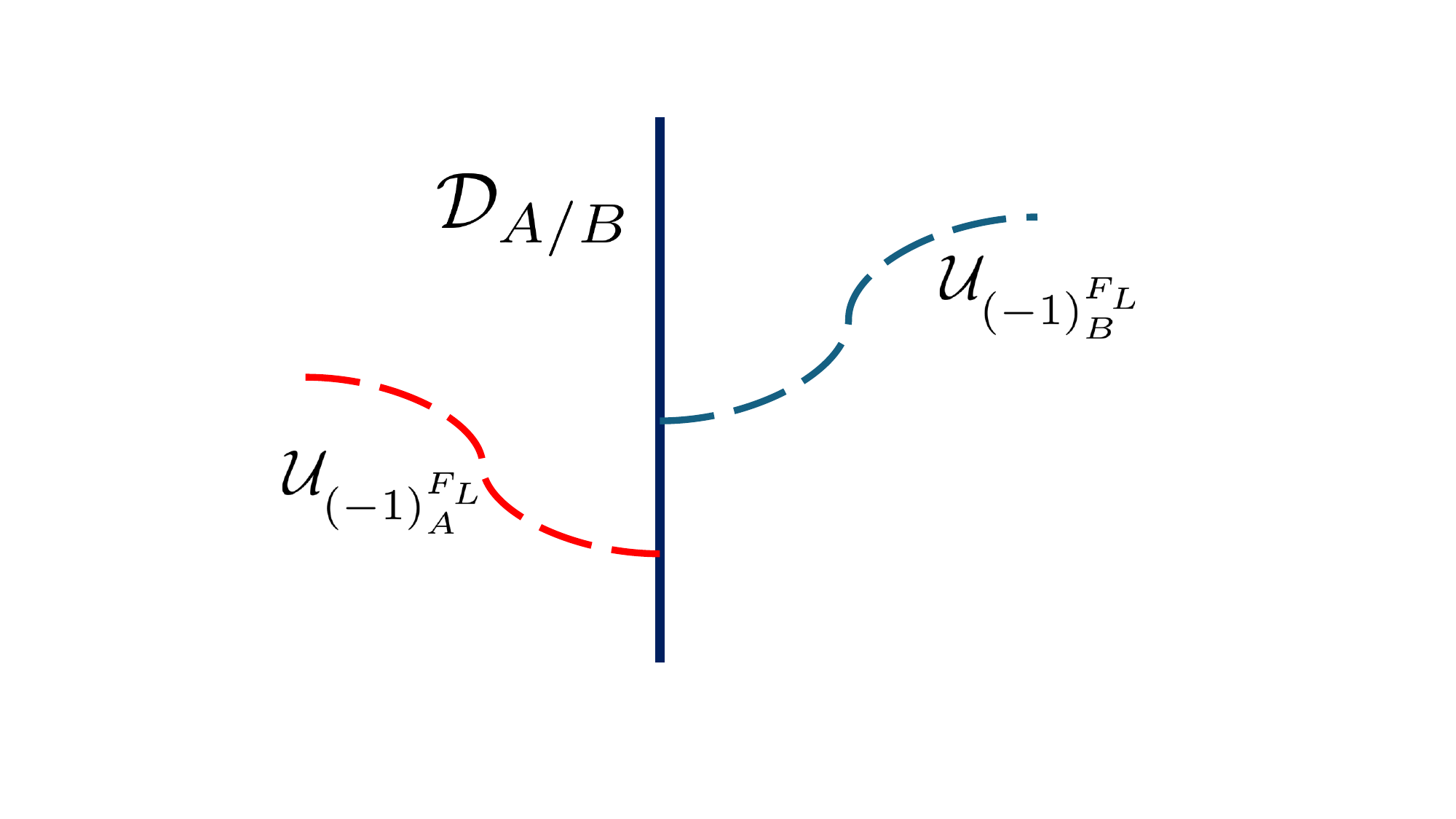}
    \caption{Endability of both of the $(-1)^{F_L}_A$ and $(-1)^{F_L}_B$ topological symmetry operators on the IIA/IIB worldsheet interface.}
    \label{fig:twoz2s}
\end{figure}

\begin{figure}[t!]
\centering
\includegraphics[trim={0 2cm 0 0cm},clip,width=15cm,scale=1]{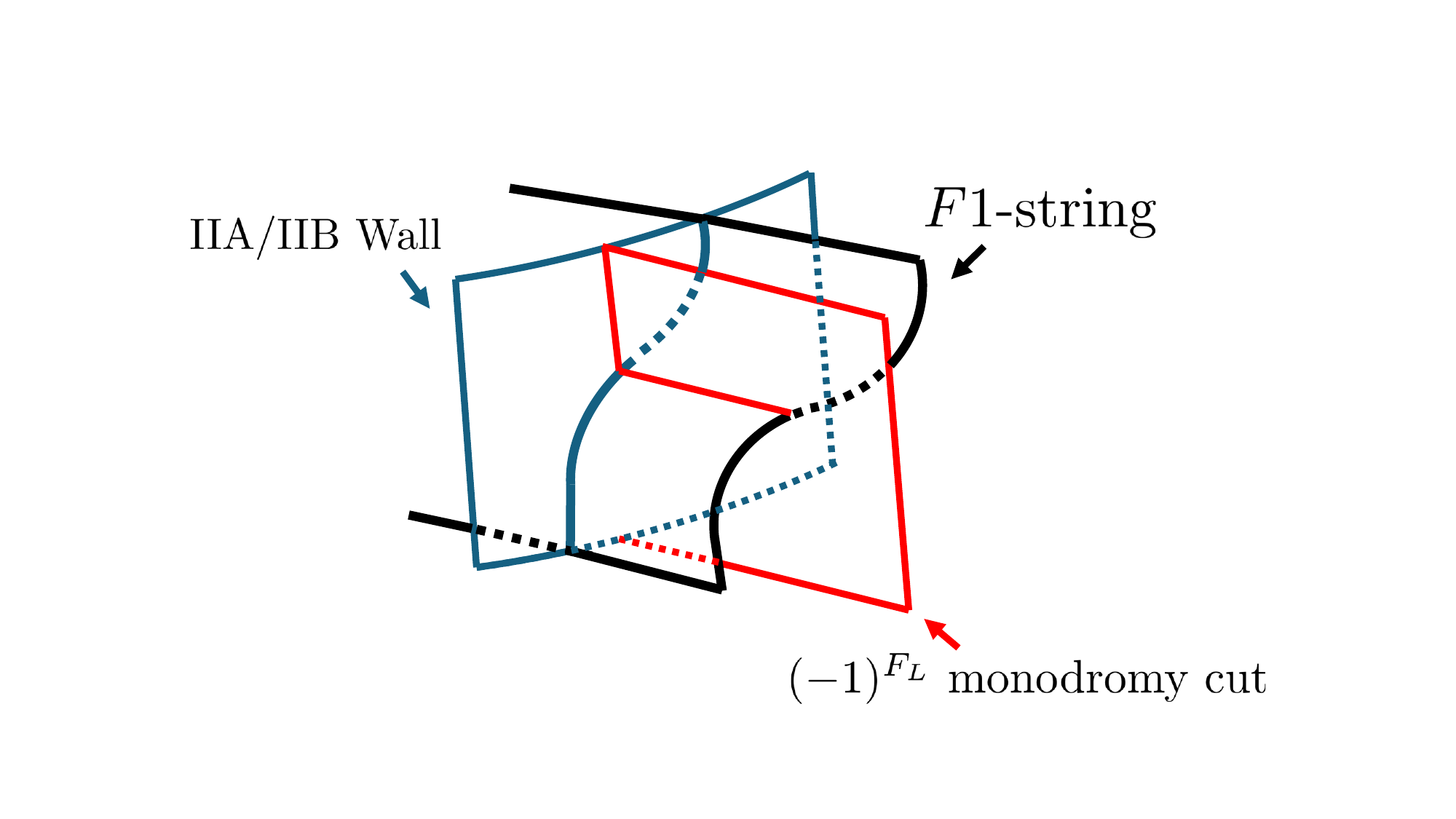}
    \caption{Depiction of a long $F1$-string in the vicinity of the IIA/IIB wall. There is a $(-1)^{F_L}$ monodromy cut which emanates out of the wall, leading to a topological operator for $(-1)^{F_L}$ ending on the interface between different GSO projections on the worldvolume of the string.} 
    \label{fig:Neumann_bc}
\end{figure}

For instance, in the Type II string, the line operators implementing the symmetry $(-1)^{F_L}$ may end on the gauging interface from either side, as depicted in Figure \ref{fig:twoz2s}. In target space, this implies that the long string stretched across the IIA/IIB wall may consistently pass through a monodromy cut for the target space gauge field $(-1)^{F_L}$. As a result, we learn that monodromy cuts for $(-1)^{F_L}$ can end on the IIA/IIB domain wall (see Figure \ref{fig:Neumann_bc}). Equivalently, the IIA/IIB domain wall serves as a Neumann boundary condition for the discrete gauge fields $(-1)^{F_L}$, so that both sides fluctuate independently.

\section{GSO Domain Walls: View from Target Space} \label{sec:TARGET}

We now turn to the target space interpretation of codimension-one GSO domain walls. As a first point, it is helpful to contrast these GSO domain walls with the $\mathrm{D}8$-brane of type IIA string theory. Recall that the $\mathrm{D}8$-brane of type IIA string theory partitions the theory into regions with differing values of the Romans mass parameter. In particular, this implies that there is a difference in the value of the cosmological constant on the two sides of such walls. This is to be expected because the IIA spectrum with Romans mass \cite{Romans:1985tz} includes a continuous RR $9$-form potential $C_9$ and an integer-quantized on-shell flux $F^{\mathrm{IIA}}_{10}$ \cite{Polchinski:1995mt}.

In the case of GSO defects, the situation is qualitatively different. By definition, the GSO domain walls source the $\mathbb{Z}_2$-valued target space parameter that controls the choice of GSO projection on the string worldsheet. In Type II string theory, this is the $\mathbb{Z}_2$ parameter considered in \cite{Distler:2009ri}, which controls whether the Type II string is Type IIA or Type IIB. In the Het$_\mathfrak{so}$ string, this  (unphysical) discrete parameter is S-dual to the vev of $C_0$ mod $2\pi$ in the Type I description, which controls the nominal chirality of the gauge group ${\rm Spin}(32)/\mathbb{Z}_2$. These torsion parameters do not correspond to any shift in the vacuum energy, and so we may take the cosmological constant to be zero on both sides of the wall. For BPS domain walls, this would force the domain wall to have vanishing tension. However, as the GSO domain walls are non-BPS, we expect them to have some positive tension associated with the potential barrier between the vacua on either side. This will prove especially important in Section \ref{sec:VORTICES}, where we consider cylindrical bubble-like configurations, since there will be no macroscopic bulk energy density which can obstruct their collapse.\footnote{Microscopically there can be a non-zero Casimir energy, but this is only expected to play a role at or below the string scale. In a long cylindrical configuration this is a subdominant effect.}

Our aim in the remainder of this section will be to establish some further properties of these GSO domain walls. We begin by studying the IIA/IIB wall from the perspective of supergravity, focusing on the structure of RR potentials on the two sides of the wall. Then, we turn to D-branes near the IIA/IIB wall. Finally, we discuss the GSO walls of Het$_{\mathfrak{so}}$ and their S-dual counterparts in type I string theory.

\subsection{RR Potentials and the IIA/IIB Wall} \label{ssec:RR}

We now discuss in greater detail the properties of the bulk IIA
and IIB\ theories in the presence of the IIA/IIB\ wall. Our discussion will be at the level of bottom-up effective field theory away from the wall, where we remain agnostic about possible boundary conditions/dynamics localized near its core. The main outcome of our discussion will be to motivate the statement that the massless RR potentials on one side of the wall are related to fluxes on the other side of the wall, namely:
\begin{equation}\label{eq:mod2relation}
C_{2k-1} = \frac{1}{2} F_{2k-1} \,\,\, \mathrm{mod} \, 2\pi  \,\,\, \text{(IIB)} \\\,\,\, \text{and} \,\,\, C_{2k} = \frac{1}{2}F_{2k}\,\,\, \mathrm{mod} \, 2\pi \,\,\, \text{(IIA)}. 
\end{equation} 
As we will see, the various non-BPS D-branes still carry a conserved $\mathbb{Z}_2$ charge associated with these $\mathbb{Z}_2$-valued potentials.\footnote{Considering Type IIB theory as an orbifold of IIA by $(-1)^{F_L}_{IIA}$, a perhaps unusual feature of \eqref{eq:mod2relation} is that $C_{2k}$ (before projecting out by the orbifold action) is part of the untwisted sector of the IIA worldsheet, while $F_{2k}$ is part of the twisted sector. Since one can characterize the (un)twisted sector of the orbifold theory as being odd (even) under $(-1)^{F_L}_{IIB}$,  \eqref{eq:mod2relation} posses no contradiction since this is a relation between $\mathbb{Z}_2$-valued quantities which are always even under sign changes. We thank O. Bergman for bringing this question to our attention.} In Appendix \ref{app:MOD2} we establish \eqref{eq:mod2relation} by directly analyzing the homotopy fixed point spectral sequence associated with the action of $(-1)^{F_L}$ on the RR fields. 

How can we see something like \eqref{eq:mod2relation} emerge in the low-energy effective field theory? The main point is that RR potentials of one type II theory are related to the RR fluxes of the other type II theory. As the argument presented in Appendix \ref{app:MOD2} is rather formal, our aim in the remainder of this subsection will be to motivate this connection by way of a minimal effective action which fits the requisite data.

As a first set of comments, observe that the IIA and IIB theories have a different set of massless higher-form potentials. In particular, the IIA side of the wall supports massless odd-form RR potentials, while the IIB side instead has massless even-form potential. Any putative effective field theory which includes the IIA/IIB wall must somehow keep track of both sets of RR fields, in such a way that the even-form potentials become gapped on the IIA side, while the odd-form potentials become gapped on the IIB side.

We work in a formalism where our path integral is over field strength tensors along with Lagrange multipliers to enforce the desired relation between the curvature of RR potentials and fluxes. Appendix \ref{app:LAGMULT} reviews this formulation for the standard type II supergravity theories. A simple effective action which fits all the requisite data is the hybrid Lagrangian:\footnote{As reviewed in Appendix \ref{app:LAGMULT}, in the Lagrangian for just the IIA or IIB Lagrangian we have $\mathbb{R}$-valued ``fluxes'' and $U(1)$-valued Lagrange multipliers. Since we are interchanging the roles of fluxes and Lagrange multipliers on the two sides of the wall we can take either the $K$'s or $L$'s to be $\mathbb{R}$-valued, and the other set to be $U(1)$-valued. At least on-shell, this recovers the expected statement that we get curvatures for $U(1)$ bundles with properly quantized fluxes. As far as our modest aim of motivating a relation between the RR potentials on one side and fluxes on the other, the present treatment suffices.} 
\begin{align}
L_{\text{IIA/IIB}} &  =\frac{\Theta_{\text{A}}}{2}\left( \left\vert K_{0}\right\vert
^{2}+\left\vert K_{2}\right\vert ^{2}+\left\vert K_{4}\right\vert ^{2}\right)
+\frac{\Theta_{\text{B}}}{2}\left(  \left\vert L_{5}\right\vert ^{2}+\left\vert
L_{7}\right\vert ^{2}+\left\vert L_{9}\right\vert ^{2}\right)
\label{eq:HybridOne}\\
&  -K_{4}\wedge dL_{5}-K_{2}\wedge d L_{7}-K_{0}\wedge dL_{9}%
,\label{eq:HybridTwo}%
\end{align}
where in the above we have introduced indicator functions $\Theta_{A}$ and $\Theta_{B}$ which are unity on their respective sides of the wall and zero otherwise (a non-zero thickness for the wall can be introduced by using more general smoothing functions). In this Lagrangian, we have even-degree form fields $K_0, K_2, K_4$, and odd-degree form fields $L_5, L_7, L_9$, which will play complementary roles on either side of the wall. There can in principle be additional degrees of freedom on the wall itself which couple to these bulk fields; if present they will produce additional localized contributions to the equations of motion, as well as the various boundary conditions. For now we remain agnostic on the presence (or absence) of such modes and couplings.\footnote{We note that in Type II closed string field theory one must similarly include additional degrees of freedom \cite{Sen:2015uaa} which ultimately decouple. As shown explicitly in \cite{Mamade:2025jbs}, this includes an additional RR-sector $p$-form fields for all $p$ with the  ``wrong" degree compared to the usual Type II RR-sector.}

Observe that the equations of motion are:%
\begin{align}
\Theta_{\text{IIA}}\ast K_{2k}-dL_{9-2k}  & =0\label{eq:Aside}\\
\Theta_{\text{IIB}}\ast L_{9-2k}-dK_{2k}  & =0,\label{eq:Bside}%
\end{align}
for $k=0,1,2$. In the IIA bulk we get the expected equations of motion under the identifications $C_{9-2k}^{\text{IIA}} = L_{9-2k}, F_{10-2k}^{\text{IIA}} = \ast K_{2k}$, in a formulation which favors the RR\ potentials $C_{9}^{\text{IIA}}$,
$C_{7}^{\text{IIA}}$, and $C_{5}^{\text{IIA}}$. In the IIB bulk we get the expected equation of motion under the identifications $C_{2k}^{\text{IIB}} = K_{2k}, F_{2k+1}^{\text{IIB}} = \ast L_{9-2k}$ in a
formulation which favors the RR\ potentials $C_{4}^{\text{IIB}}$,
$C_{2}^{\text{IIB}}$, and $C_{0}^{\text{IIB}}$.

Note that, as we cross the wall from IIA to IIB, the IIA RR potentials $C_{\text{odd}}^{\text{IIA}}$ are reinterpreted on-shell as
the IIB\ fluxes $F_{\text{odd}}^{\text{IIB}}$, while conversely the IIB potentials $C_{\text{even}}^{\text{IIB}}$ are reinterpreted on-shell as the IIA\ fluxes $F_{\text{even}}%
^{\text{IIA}}$.\footnote{Though our presentation has favored a particular electric /
magnetic split for the flux we can of course use the Hodge dual fluxes, though
this does lead to a few complications. The main issue is the treatment of the
Lagrange multiplier for the flux $F^{\mathrm{IIB}}_{5}$ and its relation to
the IIA flux $F_{4}^{\text{IIA}}$.

As a few additional (speculative) comments, one could also opt for a ``democratic'' pseudo-Lagrangian in which all fluxes and potentials appear on an equal footing (aside from the Romans mass which would require us to introduce a ``$K_{10}$'' and an ``$L_{-1}$''):
\begin{equation}
L_{\text{IIA/IIB}}^{\text{democratic}} = \underset{0 \leq k \leq 4}{\sum}
\left( \frac{\Theta_{\text{IIA}}}{2} \vert K_{2k} \vert^{2} + \frac{\Theta_{\text{IIB}}}{2} \vert L_{2k+1} \vert^2 - K_{2k} \wedge d L_{9 - 2k} \right). 
\end{equation}
This formulation is particularly suggestive since it would naturally be compatible with a tower of Stueckelberged fields in which we extend the range of $k$ to be all of $\mathbb{Z}$, in line with the appearance of a formal tower of RR potentials observed in Appendix \ref{app:MOD2}. Making sense of such a formal sum (let alone identifying the appropriate field content) is a task we leave to the interested reader.}
The hybrid action (\ref{eq:HybridOne}, \ref{eq:HybridTwo}) illustrates that the field content of IIA and IIB is such that the RR fluxes on one side of the wall can play the role of Lagrange multipliers on the other side. Moreover, we observe that this is compatible with the on-shell relation (\ref{eq:mod2relation}). In summary, this effective action motivates an interpretation in which the RR potentials on one side of the wall are related to the RR fluxes on the other. Let us emphasize that the effective action (\ref{eq:HybridOne}, \ref{eq:HybridTwo}) is meant to give a heuristic picture of the transition from IIA to IIB, and should not be trusted in the vicinity of the wall, where more involved boundary conditions are required.

\subsection{D-Branes Near the IIA/IIB Wall} \label{ssec:DBRANEPROBE}

In the previous subsection and especially Appendix \ref{app:MOD2}, we argued that the $\mathbb{Z}_2$ remnants of the RR potentials on one side of the IIA/IIB wall are simply the mod $2$ reduction of the RR fluxes on the other. We now consider the objects charged under these potentials and fluxes, namely D-branes. In particular, our aim will be to study D-branes which extend in a direction transverse to the wall. By Wick rotation, the same results will apply to a D-brane which is moved through the wall. Since we do not have a full construction of the IIA/IIB wall, we shall instead resort to a number of bottom-up self-consistency conditions.

\begin{figure}[t!]
\centering
\includegraphics[scale=0.55, trim = {2cm 3.5cm 0cm 0cm}]{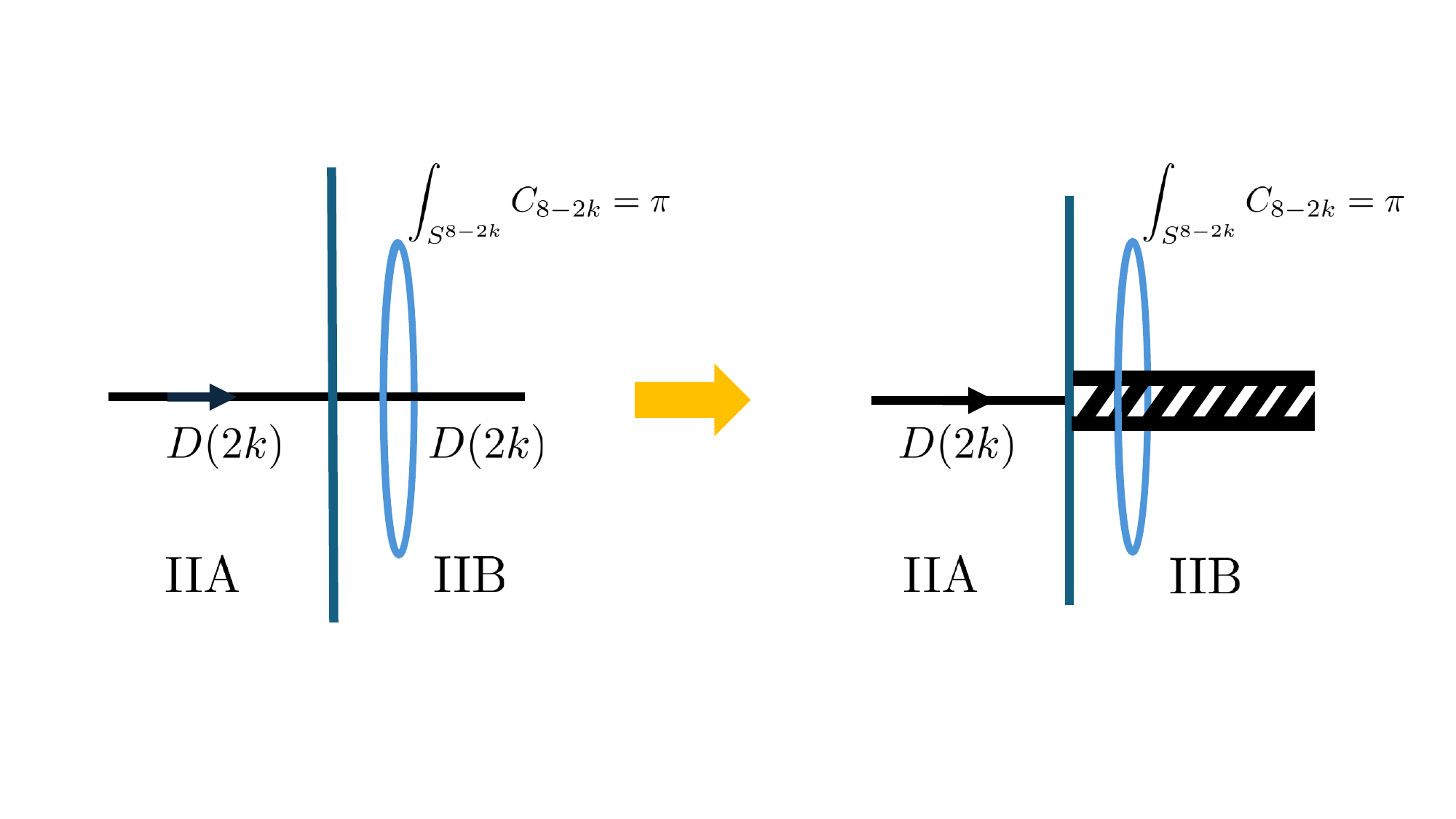}
\caption{A long D$(2k)$-brane configuration crossing the IIA/IIB wall. To the left of the IIA/IIB wall we have a BPS D-brane, while on the right we have a non-BPS D-brane. The non-BPS brane produces a holonomy $\int_S^{8-2k} C_{8 - 2k} = \pi$ on the sphere (blue circle) linking the non-BPS D-brane (see Appendix \ref{app:FLUXBRANE}). We schematically depict the decay of this non-BPS D-brane into a dilute fluxbrane.}
\label{fig:ABwall}
\end{figure}

\begin{figure}[t!]
\centering
\includegraphics[scale=0.5, trim={0 2cm 0 2cm}]{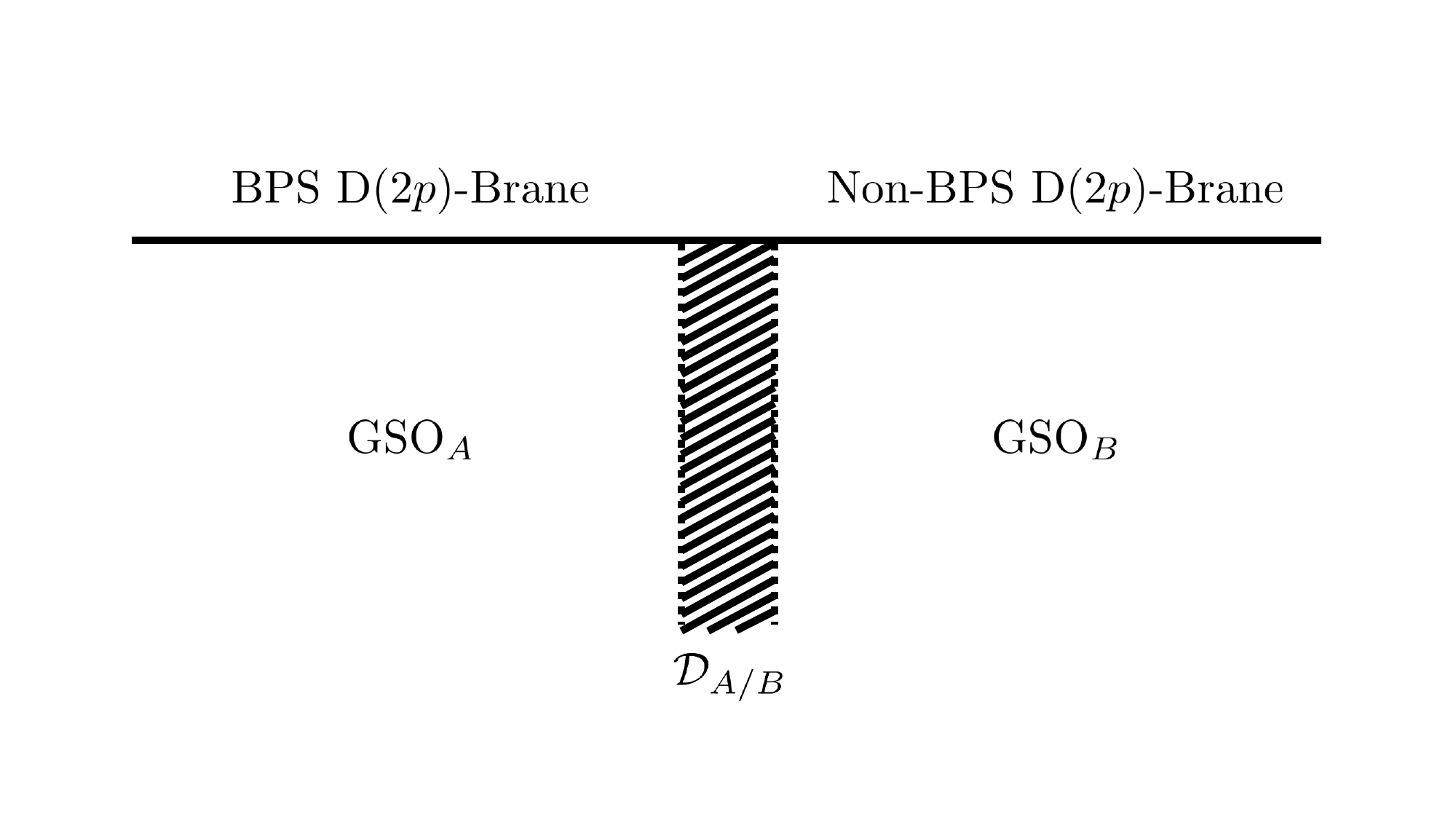}
\caption{Worldsheet depiction of a long-string endpoint attached to a $\mathrm{D}(2p)$-brane which crosses a IIA/IIB GSO wall, as characterized by the junction between the BPS boundary, the non-BPS boundary, and the bulk gauging interface $\mathcal{D}_{A/B}$ (or, whatever non-universal deformation corresponds to the IIA/IIB wall). In the target space, one endpoint of the long-string worldsheet attaches to the brane, and is moved across the wall. In the worldsheet depicted above, this is characterized by choosing a slicing where time evolution proceeds from left to right (horizontally).}
\label{fig:WorldSheetBoundaries}
\end{figure}

To proceed, consider a long D-brane which passes through the IIA/IIB wall (see Figure \ref{fig:ABwall}). From the perspective of a long string ending on the long D-brane, this configuration is characterized by an open string worldsheet with a GSO interface in the middle and corresponding boundary conditions for the D-brane on the two sides. One can also consider taking a single endpoint of the string and dragging it along the D-brane, going from one side of the wall to the other. See Figure \ref{fig:WorldSheetBoundaries} for a depiction of this process from the perspective of the worldsheet. To understand which D-branes are allowed in this configuration, we note that for each even (resp. odd) dimension $p$, there is a topological junction between the gauging interface for $(-1)^{F_L}$, the BPS or anti-BPS D$p$-brane boundary for IIA (resp. IIB), and the non-BPS D$p$-brane boundary for IIB (resp. IIA). This follows from the results of \cite{Sen:1999mg}, where descent relations between the various D-branes were considered.\footnote{One might object that \cite{Sen:1999mg} showed that the non-BPS D$p$-brane boundary is realized as the orbifold of the boundary for a D$p-\overline{{\rm D}p}$-brane pair, rather than a single D$p$-brane. However, \cite{Sen:1999mg} also showed that a single BPS D$p$-brane boundary is the orbifold of the non-BPS D$p$-brane boundary. In fact, these statements are entirely symmetrical, and can be invariantly described in terms of junctions between boundaries and the bulk gauging interface. The boundary condition for a D$p-\overline{{\rm D}p}$-brane pair is a non-simple boundary, and so a junction involving it can be decomposed as a direct sum of junctions with each of its simple summands.} Thus, a long string ending on a BPS D$p$-brane on one side of the wall can consistently move across the wall to end on a non-BPS D$p$-brane.

As a result, we propose that a BPS $\mathrm{D}p$-brane on one side of the IIA/IIB wall becomes a non-BPS D$p$-brane when pushed across the wall. This corresponds to a Wick rotation of Figure \ref{fig:ABwall}. We expect that there are nontrivial strong coupling dynamics in the vicinity of the wall, especially near the intersection of the D-brane with the IIA/IIB wall. Nevertheless, as above, our expectation is that the long-string EFT provides a reliable description of the long-distance and universal features of this brane configuration.

As a cross-check of this proposal, we may consider the RR fields sourced by the D$p$-branes on either side of the wall. The BPS D$p$-brane, as usual, sources a unit flux
\begin{equation}
    \int_{S^{8-p}} F_{8 - p} = 2 \pi,
\end{equation}
of the RR field $F_{8-p}$ on the sphere linking its worldvolume. By the relationship \eqref{eq:mod2relation} between the RR fluxes and potentials across the wall, we expect that the non-BPS D$p$-brane sources a nontrivial holonomy
\begin{equation}
    \int_{S^{8-p}} C_{8 - p} = \pi,
\end{equation}
of the RR potential $C_{8 - p}$, as depicted in Figure \ref{fig:ABwall}. This effect, though not well-known, is actually a feature of the non-BPS D-branes, as we derive in Appendix \ref{app:FLUXBRANE}. In particular, the non-BPS D-branes are not entirely uncharged: they carry a $\mathbb{Z}_2$ ``fluxbrane charge'' which may be measured at infinity by the Aharanov-Bohm phase experienced by a BPS D$(7-p)$-brane carried around the non-BPS D$p$-brane.

While the BPS side of the long-brane configuration is stable, the non-BPS D-brane on the other side of the wall is unstable. As such, it can decay via tachyon condensation to a configuration of the closed string fields. As we argue in Appendix \ref{app:FLUXBRANE}, the endpoint of this tachyon decay is a dilute fluxbrane configuration characterized by
\begin{equation}\label{eq:resulting_fluxbrane}
    \int_{\mathbb{R}_\perp^{9-p}} F_{9-p} = \pm \pi,
\end{equation}
as depicted on the right hand side of Figure \ref{fig:ABwall}. Thus, the ultimate fate of a BPS (or anti-BPS) D$p$-brane pushed through the IIA/IIB wall is a dilute $\pm \pi$-fluxbrane on the other side. Viewed in the opposite direction, this supports a picture where the BPS D-branes of IIA (resp. IIB) are merely thin, solitonic flux tubes of the RR form fields of IIB (resp. IIA), which become gapped when they cross the IIA/IIB wall.

In this context, a natural guess is that the sign $\pm \pi$ in \eqref{eq:resulting_fluxbrane} is determined by the charge of the D$p$-brane, i.e., whether it is BPS or anti-BPS. It is worth noting that, from the perspective of the long-string EFT, the charge of the D$p$-brane is determined by including or omitting a $(-1)^{F_L}$ topological line along the D-brane boundary \cite{Witten:2023snr}. It is not yet clear to us how this condition might be related to the target space physics we have been discussing.

This appears to be a relatively self-consistent picture for various D-brane configurations near the IIA/IIB wall, but there is clearly more work to be done. For example, a particularly difficult issue concerns the choice of boundary conditions for the RR potentials and fluxes along the wall. Observe that our characterization of long D-branes crossing/passing through the IIA/IIB wall treated the D$p$-branes on equal footing with their electromagnetically dual D$(6-p)$-brane counterparts. This would seem to suggest we should impose the same boundary conditions (say, Dirichlet or Neumann) for both $F_{p+2}$ and $\ast F_{p+2} = F_{8-p}$ along the wall. However, this is not consistent, as it would over-constrain the bulk dynamics.\footnote{To appreciate the puzzle from a complementary perspective, compare this with the case of a standard superconductor. While electric charges are screened, magnetic charges are confined.} Particularly challenging is the self-dual field $F_5$ of IIB, whose boundary condition necessarily involves the fermions and gravitini which cancel the gravitational anomaly of $F_5$. A final, related issue has to do with the conservation of D-brane charge across the IIA/IIB wall, as we have related the integer conserved D$p$-brane charge to the fluxbrane charge $\eta = \pi$ of the non-BPS D-brane, which is only conserved mod 2.

\subsection{The Het$_{\mathfrak{so}}$ Wall and the Type I D8-brane}

Let us now turn to the GSO wall of Het$_{\mathfrak{so}}$. In this case, the difference between the two sides of the wall is milder, since the main change is the nominal global form of the gauge group, i.e., we have a domain wall between the $\mathrm{Spin}(32) / \mathbb{Z}_{2}^{c}$ and $\mathrm{Spin}(32) / \mathbb{Z}_{2}^{s}$ gauge theories, coupled to gravity.

In fact, the S-dual of this GSO domain wall is well understood: it is precisely the non-BPS $\mathrm{D}8$-brane of Type I string theory! Indeed, as discussed in Section \ref{sec:Type_I_review}, the IIB RR-scalar $C_0$ is projected in Type I to a $\mathbb{Z}_2$-valued scalar $C_0=0 \; \mathrm{or}\; \pi \; \mathrm{mod}\; 2\pi$, which controls the nominal form of the gauge group.\footnote{As also discussed in Section \ref{sec:Type_I_review}, the precise value of $C_0$ is unphysical as it can be shifted by an anomalous $\mathrm{det}=-1$ reflection element in a (Pin-lift of) $O(32)$, as shown in \cite{BergmanSeminar, Montero:2022vva}.} As argued in \cite{Witten:1998cd, Gukov:1999yn, Montero:2022vva}, the non-BPS $\mathrm{D}8$-brane of Type I produces a jump $\Delta C_0=\pi$ across it, and so the non-BPS D8-brane of Type I may be identified as the S-dual of the GSO domain wall of Het$_{\mathfrak{so}}$. This effect is an example of our conclusions in Appendix \ref{app:FLUXBRANE}, where we argue that the non-BPS D-branes produce a $\pi$ holonomy in the RR potentials around their worldvolumes. Importantly, this 8-brane is not a GSO domain wall in the Type I description, as it is associated to the GSO projection on the heterotic string worldsheet, not the Type I worldsheet.

We can probe the Het$_{\mathfrak{so}}$ GSO wall using the non-BPS particle in the spinor representation of the gauge group, which is realized in the S-dual Type I description as the non-BPS D0-brane \cite{Sen:1998rg}. In the Het$_{\mathfrak{so}}$ description, this particle is merely a closed string state charged under the quantum symmetry $(-1)^s$ of the GSO projection. Thus, this particle is analogous to a quanta of the RR fields of Type II string theory. As these fields become gapped when brought across the IIA/IIB wall, our expectation is that the spinor cannot freely cross the GSO wall.

\begin{figure}
    \centering
    \includegraphics[width=0.7\linewidth]{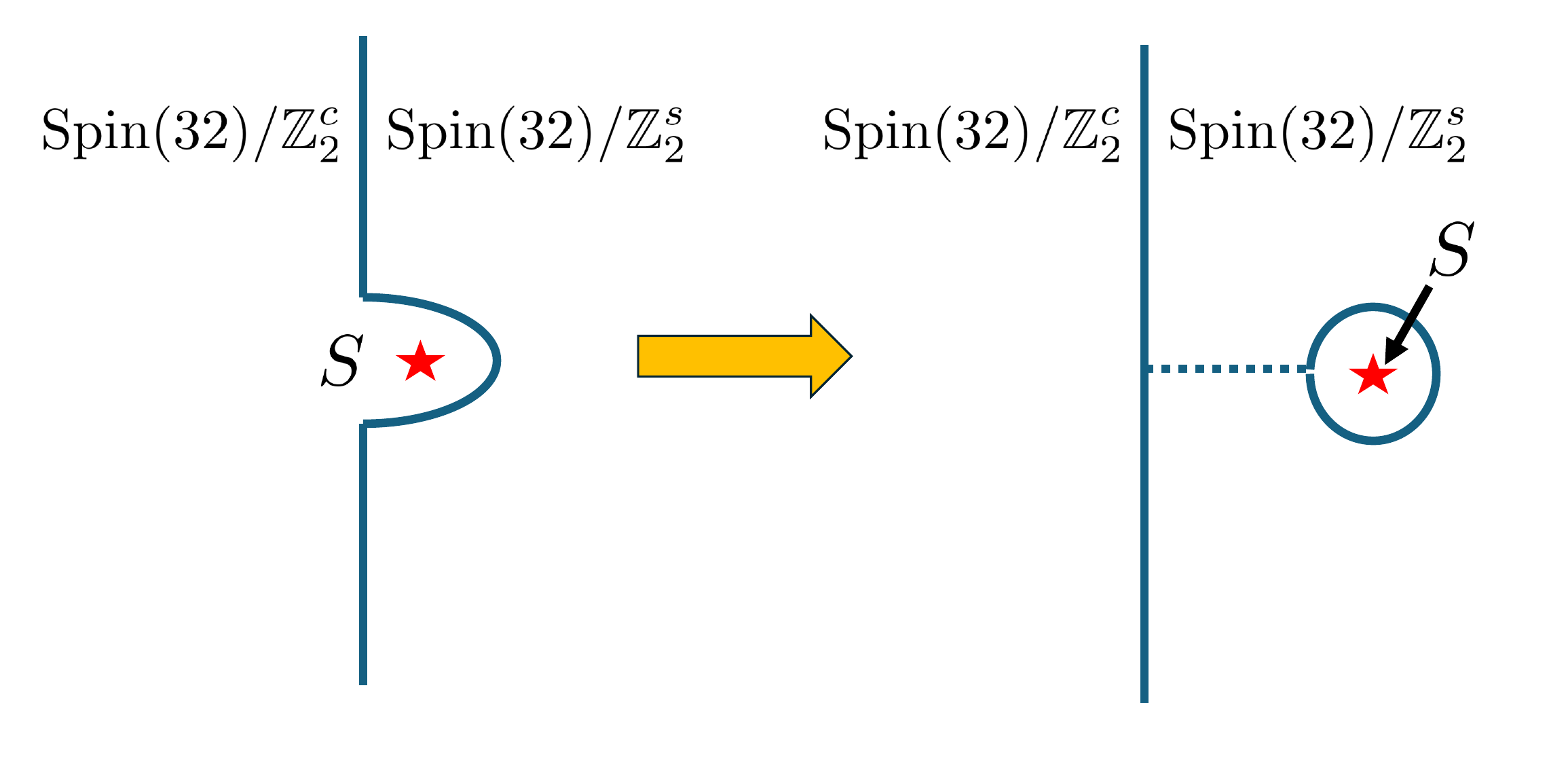}
    \caption{A spinor state in $\mathrm{Spin}(32)/\mathbb{Z}^c_2$ heterotic (or Type I) transitioning across the GSO wall (Type I $\mathrm{D}8$-brane). The end result is a particle attached to a tensionful string stretching back to the GSO wall. In Type I string theory, this is likely given by a single fundamental Type I string, analogous to the Hanany-Witten effect for pushing a Type IIA D0-brane across a D8-brane.}
    \label{fig:spinortransition}
\end{figure}

Thus, consider pushing the spinor through the GSO wall of Het$_{\mathfrak{so}}$. On the other side, the gauge group is $\mathrm{Spin}(32) / \mathbb{Z}_{2}^{s}$, which forbids the spinor representation. As depicted in Figure \ref{fig:spinortransition}, a natural guess is that the resulting particle on the other side of the wall is attached back to the wall by a tensionful string. Dynamically, this tension would exert a force on the particle, ensuring that it gets pulled back through the wall to the side where it is an allowed excitation. In the Type I description, this can be realized as a non-BPS $\mathrm{D}0$-brane attached to F1 string, in line with the known Hanany-Witten transition \cite{Hanany:1996ie} in the Type IIA D0-D8 system \cite{Billo:1998vr, Lu:2009gm, Dibitetto:2018gbk}. To reach Type I, one must orbifold this Hanany-Witten transition first by $(-1)^{F_L}$ to reach Type IIB and then by $\Omega$ to reach Type I. 

More explicitly, note that the endpoint of a Type I F1-string carries Chan-Paton factors charged in the vector representation of $\mathfrak{so}(32)$. If we stretch an F1 string deep into the $\mathrm{Spin}(32)/\mathbb{Z}^s_2$ side, we can consistently attach a $\mathrm{D}0$-brane in the spinor representation $S$ for the simple reason that the product representation $V\otimes S$ is invariant under $\mathbb{Z}^s_2$ and thus survives as a consistent state. 

\section{GSO Vortices} \label{sec:VORTICES}

Our discussion so far has focused on the codimension-one GSO walls, defined by a jump in the choice of GSO projection. In this section we turn to the codimension-two GSO vortices. As described in the Introduction, these are vortices (or ``cosmic strings'') for the $\mathbb{Z}_2$ gauge symmetry in target space associated with the quantum symmetry of the GSO projection. As the electrically charged states under this $\mathbb{Z}_2$ gauge symmetry are perturbative string states in the Ramond sector, the GSO vortices are defined by the phase experienced by a Ramond-sector string state when moved around the vortex. The existence of GSO vortices is predicted by the Completeness Hypothesis \cite{Polchinski:1995mt, Banks:2010zn, Heidenreich:2021xpr}, or equivalently, by the Cobordism Conjecture \cite{McNamara:2019rup}. We focus our attention on two examples of GSO vortices: the $F_L$ R7 brane in type II and the non-BPS 7-brane in $\text{Het}_{\mathfrak{so}}$ which is S-dual to the type I D7-brane. Many properties of these objects have been described at length elsewhere \cite{Witten:1998cd, Frau:1999qs,  Debray:2023yrs, Dierigl:2023jdp}. Our plan in this section will be to give a topological characterization of GSO vortices and relate them to the collapse of GSO walls wrapped in a cylindrical configuration. We defer issues concerning the stability and dynamics of such configurations to Section \ref{sec:STABILITY}.

\subsection{$F_L$ $\mathrm{R}7$-branes in Type II}

We first turn to the GSO vortices of the type II theories. In Type IIB, this is the $F_L$ $\mathrm{R}7$-brane discussed in references \cite{Dierigl:2022reg, Debray:2023yrs}, whose monodromy involves reflection on a cycle of the F-theory torus. In Type IIA, we have an analogous $F_L$ $\mathrm{R}7$-brane, whose monodromy corresponds to a reflection of the M-theory circle. Lifting this configuration to M-theory produces the cobordism defect trivializing the class of M-theory on a Klein bottle, another object predicted by the Cobordism Conjecture \cite{McNamara:2019rup}.

Many properties of the IIB $F_L$ $\mathrm{R}7$-brane were established in \cite{Dierigl:2022reg,Debray:2023yrs, Dierigl:2023jdp}. Monodromy around the $F_L$ $\mathrm{R}7$-brane causes all of the RR potentials to flip sign, i.e., $C_{\mathrm{even}} \mapsto - C_{\mathrm{even}}$, so that it acts as a generalization of an Alice string (see \cite{Schwarz:1982ec}). By considering lasso-type configurations in which BPS D-branes wind around the $\mathrm{R}7$-brane, \cite{Dierigl:2022reg} argued that a pair of D-branes can terminate on the GSO vortex. In order to maintain gauge invariance $C_{2k} \mapsto C_{2p} + d \lambda_{2p -1}$, this at the very least requires the GSO vortex to support odd degree potentials, i.e., $b_{2k-1}$. Indeed, the appearance of odd degree potentials on the worldvolume of the GSO vortex is strong evidence that this vortex is in fact related to our GSO wall; observe that in crossing the IIA/IIB wall, the spectrum of massless RR potentials changes, as per our discussion in Section \ref{ssec:RR} and Appendix \ref{app:MOD2}.

\begin{figure}[t!]
    \centering
\includegraphics[trim={0 1cm 0 2cm},scale=0.4]{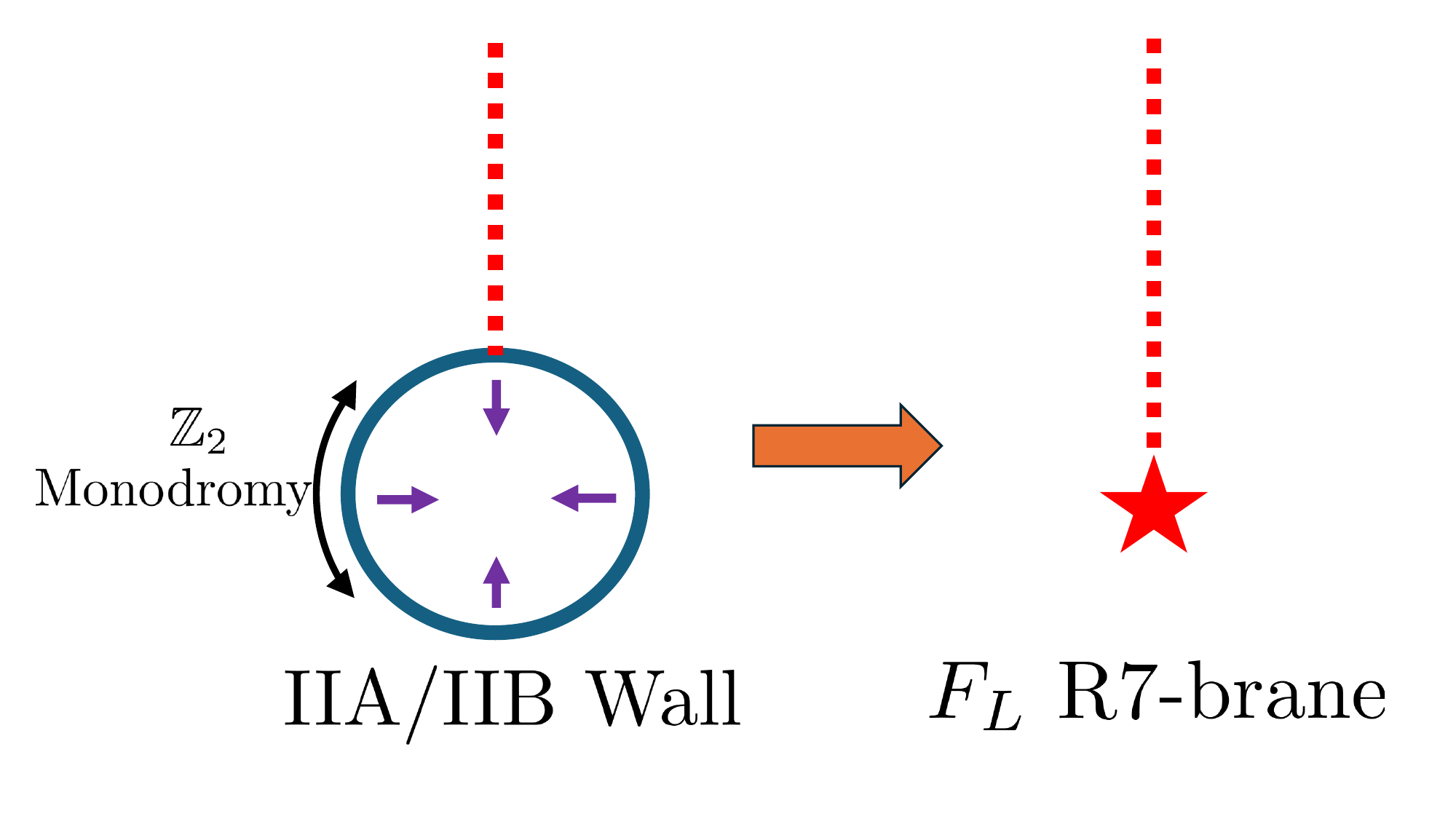}
    \caption{Depiction of the collapse of the IIA/IIB wall to the $(-1)^{F}_L$ $\mathrm{R}7$-brane of type IIB string theory (the eight dimensions parallel to the R7-brane worldvolume have not been drawn). When IIA is on the outside, one instead reaches the $(-1)^{F_L}$ $\mathrm{R}7$-brane of type IIA string theory.}
    \label{fig:Collapso}
\end{figure}

We now argue for a direct relationship between the GSO vortex to the GSO wall. Instead of having the wall fill a plane $\mathbb{R}^{8,1}\subset \mathbb{R}^{9,1}$ we can instead consider a configuration in which it is wrapped over a circle embedded in $\mathbb{R}^2$, so that the spatial profile of the wall is topologically $\mathbb{R}^7 \times S^1$ (see Figure \ref{fig:Collapso}). Recall from Section \ref{sec:LONGSTRING} that it is possible to end a $(-1)^{F_L}$ monodromy cut on a IIA/IIB wall. From the bulk perspective, this implies that there exists a boundary condition for the $(-1)^{F_L}$ bulk gauge field which can be twisted around the cycle $S^1$ in the worldvolume of the IIA/IIB wall. Putting this together, we learn that a GSO-vortex and the GSO-wall in a cylindrical configuration with a discrete monodromy have the same charge with respect to the $(-1)^{F_L}$ gauge symmetry!\footnote{What happens if we wrap the IIA/IIB wall on a general $S^{n}$? To prevent collapse to nothing in the cylindrical configuration, we needed the monodromy cut to terminate on a codimension-two cycle, but none is available for lower-dimensional spheres. There are almost certainly other objects which may be dissolved in the IIA/IIB wall, but our focus here is on the R7-brane.} Put differently, we see that the endpoint of a $(-1)^{F_L}$ monodromy cut on the IIA/IIB wall can be identified with a dissolved R7-brane. Due to the tension of the IIA/IIB wall, this cylindrical configuration will tend to shrink, presumably stabilizing at some small radius due to quantum mechanical effects. We identify the end point of this process as the R7-brane itself.

\begin{figure}[t!]
    \centering
\includegraphics[trim={0 2cm 0 2cm},clip,width=16cm,scale=0.8]{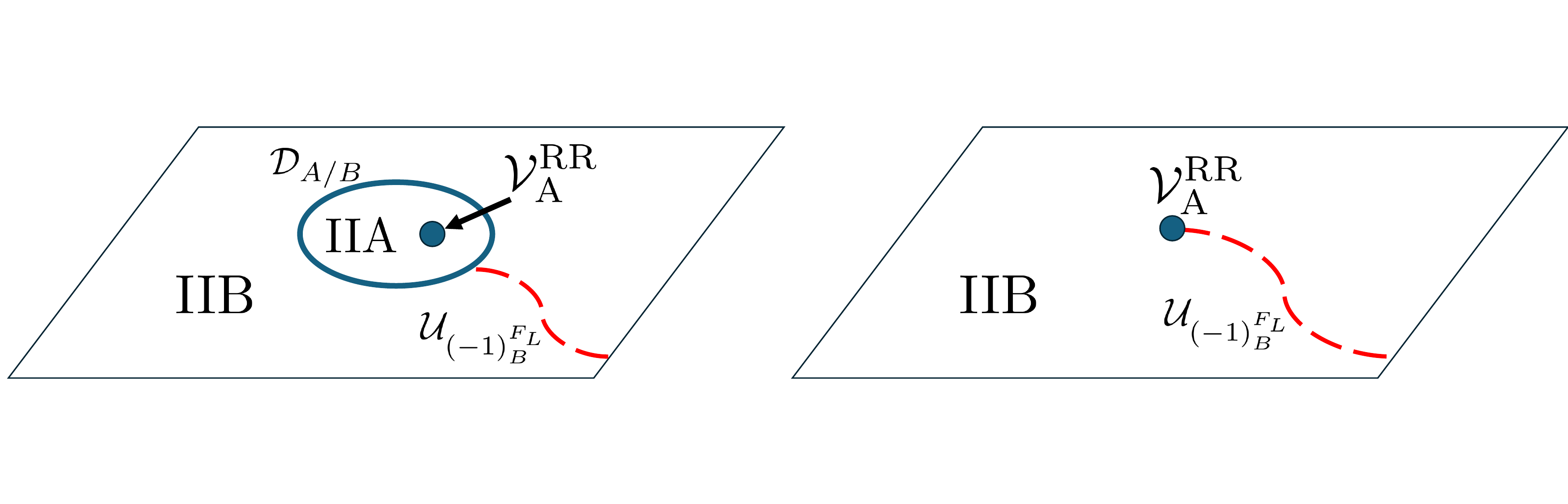}
    \caption{On the left, we depict a Euclidean worldsheet which includes a IIA/IIB interface, now wrapped on a compact region. There is a $(-1)^{F_L}_B$ monodromy around a circle in its worldvolume. This worldsheet embeds in the target space as a transverse spatial slice of Figure \ref{fig:ABwallIntro}. On the right we show the collapse of the interface region which, in the target space, corresponds to the collapse of the cylindrical IIA/IIB wall configuration along a contractible circle. On the worldsheet we must terminate the $(-1)^{F_L}_B$ symmetry operator on a local operator which does not survive the IIB GSO projection but \textit{does} survive the IIA GSO projection. These are given precisely by the RR-sector vertex operators of IIA. Note that this discussion should be understood in the context of the long-string EFT, as there are difficulties in extending it to the case of the critical string \cite{WorlsheetVortices}.}
    \label{fig:abr7}
\end{figure}

In particular, observe that in the limit where the bubble is very small, with IIB on the outside, the zero modes of the bulk IIA RR potentials exactly match the spectrum of potentials $b_{\mathrm{odd}}$ expected on a GSO vortex. This can also be understood from the worldsheet EFT of the long string, as the collapse simply sends vertex operators for the IIA RR fields to the twisted sectors of $(-1)^{F_L}$ on the IIB side, which correspond to the IIB RR fields.\footnote{There are difficulties in making this picture consistent for critical strings \cite{WorlsheetVortices}.}
This is depicted in Figure \ref{fig:abr7}.
Moreover, since we also know that a \textit{single} D$({\rm odd})$-brane of type IIB transmutates to a fluxbrane on the other side of the IIA/IIB wall, this relation between GSO vortices and GSO walls also establishes that a single D-brane can terminate on a GSO vortex. As a consequence, the charge lattice for the $b_{2k-1}$-potentials on the $F_L$ $\mathrm{R}7$-brane populates $\mathbb{Z}$.\footnote{In \cite{Dierigl:2022reg} a conservative interpretation of lasso-type configurations implies that an even number of branes can terminate, i.e., one would only get a $2 \mathbb{Z}$ lattice. 
For an odd number of branes this requires brane-junction configurations which include fractionally charged constituents. That being said, such fractional string junction configurations are known to be sensible in certain supersymmetric contexts, and provide a uniform characterization of higher-form charge structures for some branes (see e.g., \cite{Cvetic:2021sxm, Cvetic:2022uuu}).}

While we have focused on the case of the type IIB GSO vortex and a GSO wall which collapses to IIB on the outside, we can of course consider the closely related case of a type IIA GSO vortex and a GSO wall which collapses to IIA on the outside. In this case, the primary change is that the spectrum of RR potentials and BPS D-branes is of course different, but we learn that the IIA GSO vortex supports $U(1)$-valued $b_{\mathrm{even}}$ potentials, and that D(even)-branes can end on it.

\subsection{GSO Vortices in Het$_\mathfrak{so}$}

Let us now turn to the case of GSO walls and vortices in Het$_{\mathfrak{so}}$ string theory and its S-dual Type I string theory. As discussed in Sections \ref{sec:het_review} and \ref{sec:Type_I_review}, the quantum symmetry of GSO in Het$_{\mathfrak{so}}$ string theory is the $\mathbb{Z}_2$ center of the gauge group ${\rm Spin}(32)/\mathbb{Z}_2$, which in Type I is given by the $\mathbb{Z}_2$ $K$-theory gauge field under which the non-BPS D0-brane is charged (see Appendix \ref{app:EXTENSION}). As shown in \cite{Gukov:1999yn}, this vortex is simply the non-BPS D7-brane of Type I.

\begin{figure}[t!]
\centering
\includegraphics[scale=0.30, trim = {0cm 0cm 0cm 0cm}]{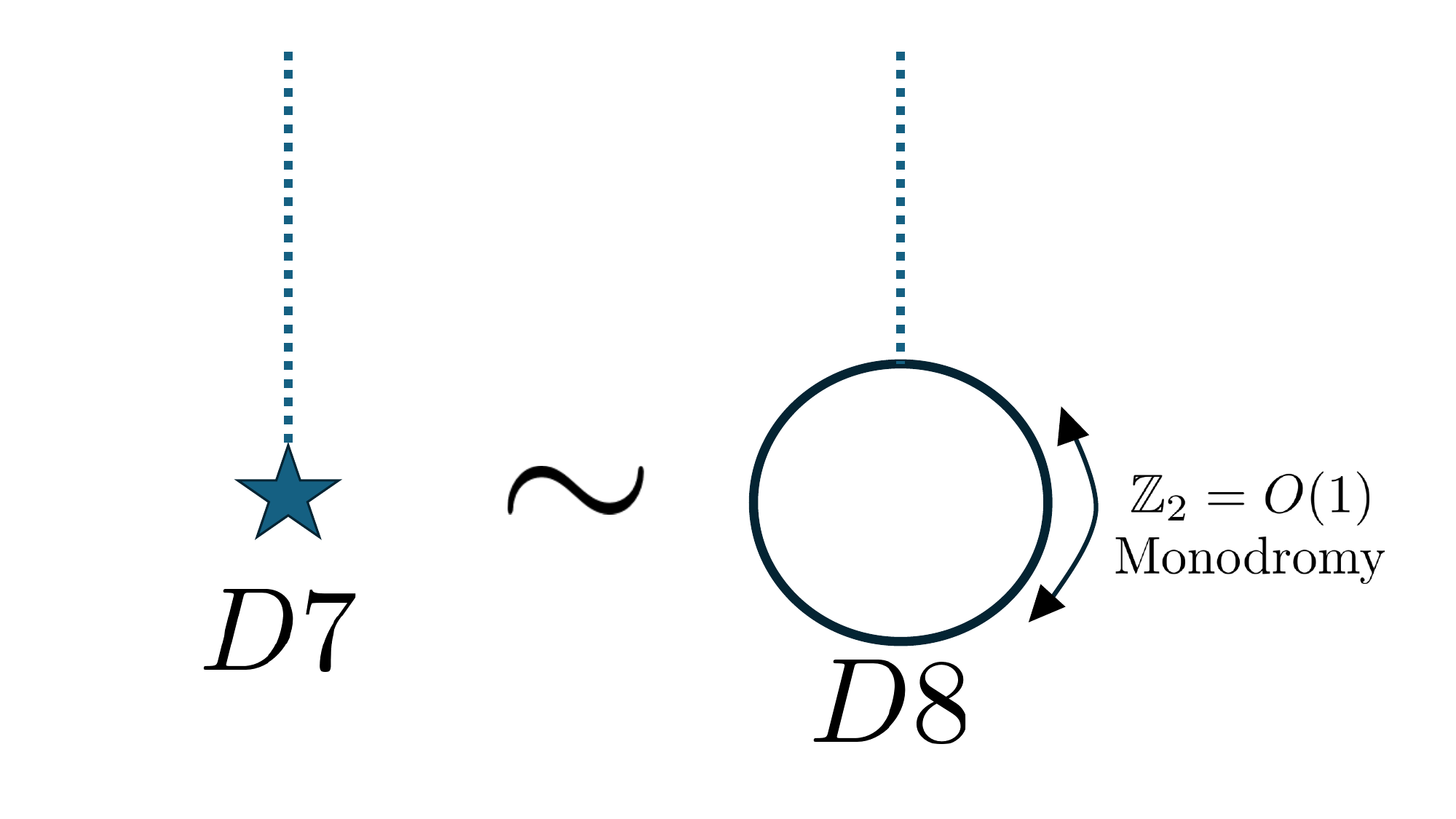}
\caption{A $\mathrm{D}7$-brane carries the same charge as a single $\mathrm{D}8$-brane with an $O(1)$ Wilson line along its $S^1$ direction in its worldvolume. This equivalence relation in $\mathrm{D}7$-brane charge is denoted by $\sim$ above.}
\label{fig:D7explosion}
\end{figure}

Applying the same reasoning as above, we expect that the GSO vortex of Het$_{\mathfrak{so}}$ can equally well be constructed by starting off-shell with a GSO domain wall wrapped on a circle with a discrete homolomy, and contracting it to a small size. In Type I, this corresponds to an off-shell circular configuration of a D8-brane contracting to a D7-brane, as in Figure \ref{fig:D7explosion}.\footnote{On shell, the situation is considerably more delicate because the wall and the vortex may suffer from instabilities. Indeed, in Section \ref{sec:STABILITY} we argue that the type II GSO defects are stable, whereas the Het$_{\mathfrak{so}}$ GSO defects are unstable.} We confirm this in Type I language in Appendix \ref{app:TYPEI} through an analysis of non-BPS D-brane configurations.

\section{Stability Considerations} \label{sec:STABILITY}

Our discussion so far has focused on topological properties of the GSO defects. In this section we turn to the stability of these objects. We begin with an analysis of GSO vortices, which we have identified with a collapsed GSO wall wrapped in a cylindrical configuration. We then return to stability of the GSO walls themselves.

We now argue---contrary to previous expectations---that the $(-1)^{F_L}$ $\mathrm{R}7$-branes of the Type II theories are actually stable, while the GSO vortices of Het$_{\mathfrak{so}}$ are unstable (in accordance with the known instabilities of their S-duals). The main criterion we use to evaluate stability is whether there is a smooth field configuration in the bulk which allows these localized configurations to expand out in size. Indeed, this criteria was used in \cite{Kaidi:2024cbx} to argue for the stability of new non-BPS branes in heterotic string theory. Said differently, we know that these GSO vortices carry a $\mathbb{Z}_2$ magnetic gauge charge, and we are asking whether this gauge charge can be realized by a smooth configuration of lower energy density. If the $\mathbb{Z}_2$ gauge group embeds in a connected, continuous, internal gauge group, we can realize the asymptotic holonomy of the GSO vortex by a smooth gauge field configuration with nonzero magnetic flux.\footnote{That the gauge fluxes of a continuous, connected gauge group provide a complete spectrum of vortices was argued in \cite{Heidenreich:2021xpr}.} In the case of the IIA/IIB wall there is no such configuration, since $(-1)^{F_L}$ is a discrete gauge symmetry of the Type II theories. On the other hand, the $\mathbb{Z}_2$ holonomy of the Het$_{\mathfrak{so}}$ GSO vortex embeds in the center of $\mathrm{Spin}(32)/\mathbb{Z}_2$. As such, we expect that the GSO vortex of type II theories is stable, whereas we expect the GSO defect of Het$_{\mathfrak{so}}$ (and the S-dual $\mathrm{D}7$-brane of Type I) to be dynamically unstable, puffing up into a flux configuration.

Could there be some other quasi-singular configurations which the $\mathrm{R}7$-branes of the Type II theories decay into? While the absence of a microscopic characterization obstructs a full analysis, we find such a possibility implausible. For instance, consider a configuration in which the R7-brane puffs up into some cylindical domain wall, analogous to the cylindrical IIA/IIB wall considered above. Due to the tension of the domain wall, this configuration will only have a chance to expand if the vacuum energy is lower on the inside than the outside of the wall. In particular, one might imagine a domain wall separating the Type II vacuum with some putative AdS$_{10}$ vacuum. However, such a configuration would in turn sow the seeds for an instability of the Type II vacuum itself, contradicting its known stability. Even worse, the gravitational attraction of the domain wall would ensure that sufficiently large bubbles would be hidden behind event horizons, as was recently discussed in \cite{Sen:2025oeq}.

It is worthwhile to contrast this vortex with seemingly similar codimension-two 
defects which arise from non-supersymmetric orbifolds such as $\mathbb{C} / \mathbb{Z}_n$ (see \cite{Adams:2001sv}).\footnote{One might try to relate the type II GSO vortex to a generalized orbifold for the quantum $(-1)^{F_L}$ combined with a spacetime group action. This is destined to fail, however, since the GSO vortex is intrinsically strongly coupled \cite{Dierigl:2022reg, Sappho, WorlsheetVortices}. See also \cite{Heckman:2024zdo} for a recent discussion on how introducing probe D3-branes can potentially stabilize related non-supersymmetric orbifolds.} In that case, an initially localized tachyon in a closed string twisted sector condenses, leading to an eventual smoothing out of the singularity. Said differently, at late times/long distance scales the initial localized instability is eventually detected in terms of a dissipating smooth field configuration which can be consistently described in the gravitational effective field theory. In the case of the type II GSO vortices there is no such candidate ``smoothed out configuration.''

We now discuss the stability of the GSO walls themselves. The same supergravity considerations used to establish stability of the type II GSO vortices apply equally well to the GSO walls: there is no smooth supergravity configuration providing a domain wall between IIA and IIB. By contrast, the Het$_{\mathfrak{so}}$ GSO wall is unstable, as there is a smooth configuration connecting the two nominal forms of $\mathrm{Spin}(32)/\mathbb{Z}_2$: a transition function for the anomalous reflection symmetries in the disconnected component of (a Pin lift of) $\mathrm{O}(32)$ \cite{BergmanSeminar, Montero:2022vva}. The decay of the Het$_{\mathfrak{so}}$ GSO wall is an example of the instability of domain walls for spontaneously broken $\mathbb{Z}_2$ gauge symmetries due to nucleation of ``cosmic strings'' (see, e.g., \cite{Kibble:1982ae, Vilenkin:1984ib, Kibble:1982dd, Everett:1982nm}). In the Type I language, this instability is due to an open string tachyon in the $8-9$ sector of the $\mathrm{D}8$-brane as discussed in Appendix \ref{app:TYPEI}.

\section{Conclusions} \label{sec:CONC}

The Swampland Cobordism Conjecture predicts the existence of new objects in quantum gravity. In this paper we have taken this conjecture at face value, and given worldsheet and target space characterizations of the predicted codimension-one and codimension-two defects associated with the GSO projection. In particular, we have seen that the GSO interface in the long-string EFT captures many properties of the GSO walls in Type II and Het$_{\mathfrak{so}}$ string theory. Moreover, this same characterization can be used to provide supporting evidence for the predicted codimension-two GSO vortices, including, in particular, the R7-branes of Type IIA and IIB. By tracking cylindrical configurations of these same walls, and studying related field configurations, we argued for the stability of the $F_L$ $\mathrm{R}7$-brane of type II theories, as well as confirmed the instability of the Het$_{\mathfrak{so}}$ S-dual of the Type I D7-brane.

Having established strong evidence for the existence of GSO walls, GSO vortices, and the relationship between the two, it is natural to ask for further information on their properties. For example, it would be very interesting to extract the tension of these objects, as well as further features of the corresponding worldvolume theories.\footnote{One might also attempt to use holography to extract the tension of these GSO defects. See e.g., \cite{Heckman:2025isn} for some computations along these lines.} A first step in this direction would be constructing black brane supergravity solutions corresponding to the R7-branes, which should exist given our arguments for their stability. With such a solution in hand, one could find a worldsheet CFT describing strings moving in its near-horizon region, as was done for non-BPS heterotic branes in \cite{Kaidi:2023tqo, Kaidi:2024cbx}. One might also attempt to construct a solution corresponding to the GSO wall, though this appears very challenging as it would likely require understanding the boundary conditions of the chiral fields in Type IIB.

A rather striking outcome of our analysis is the stability of the $\mathrm{R}7$-brane of type IIB string theory. Indeed, since we have established the stability of one such brane, it is natural to expect that all objects in the same IIB duality orbit, such as  the $\Omega$ R7-brane (the S-dual of the $F_L$ R7-brane \cite{Dierigl:2022reg}), are also stable. Going the other direction, this motivates a further natural question: is there an off-shell configuration where the 
$\Omega$ R7-brane has puffed up to finite size as a cylindrical configuration of some other domain wall? 
Presumably, the long-string characterization will involve placing an orientation-reversal defect on the worldsheet. It would be very interesting to provide further evidence for the existence of this and related codimension-one walls.

Supercritical string theories have been used to produce an intricate web of transitions between different perturbative string theories \cite{Hellerman:2004qa, Hellerman:2006nx, Hellerman:2007fc, Hellerman:2007ym}. Given this, it is natural to ask how the considerations of the present work fit with these constructions. One possibility is that the vicinity of a GSO defect may be describable in terms of a supercritical string theory. We leave an exploration of this intriguing possibility to future work.

What is the microscopic physics characterizing the IIA/IIB transition? As a speculative option, since we have seen that RR potentials on one side of the wall leave behind a $\mathbb{Z}_2$ remnant associated with fluxes on the other side of the wall, it is tempting to imagine a BCS-type theory of superconducitivity for RR potentials. This would suggest a picture in which the IIA/IIB transition involves condensation of ``Cooper pairs'' of D-branes in order to break the continuous RR gauge fields to $\mathbb{Z}_2$ subgroups. An issue with this picture, related to the discussion of boundary conditions in Section \ref{ssec:DBRANEPROBE}, is that it would require simultaneous condensation of both electric and magnetic charges under the RR gauge fields.

\begin{figure}[t!]
\centering
\includegraphics[scale=0.40, trim = {0cm 2cm 0cm 0cm}]{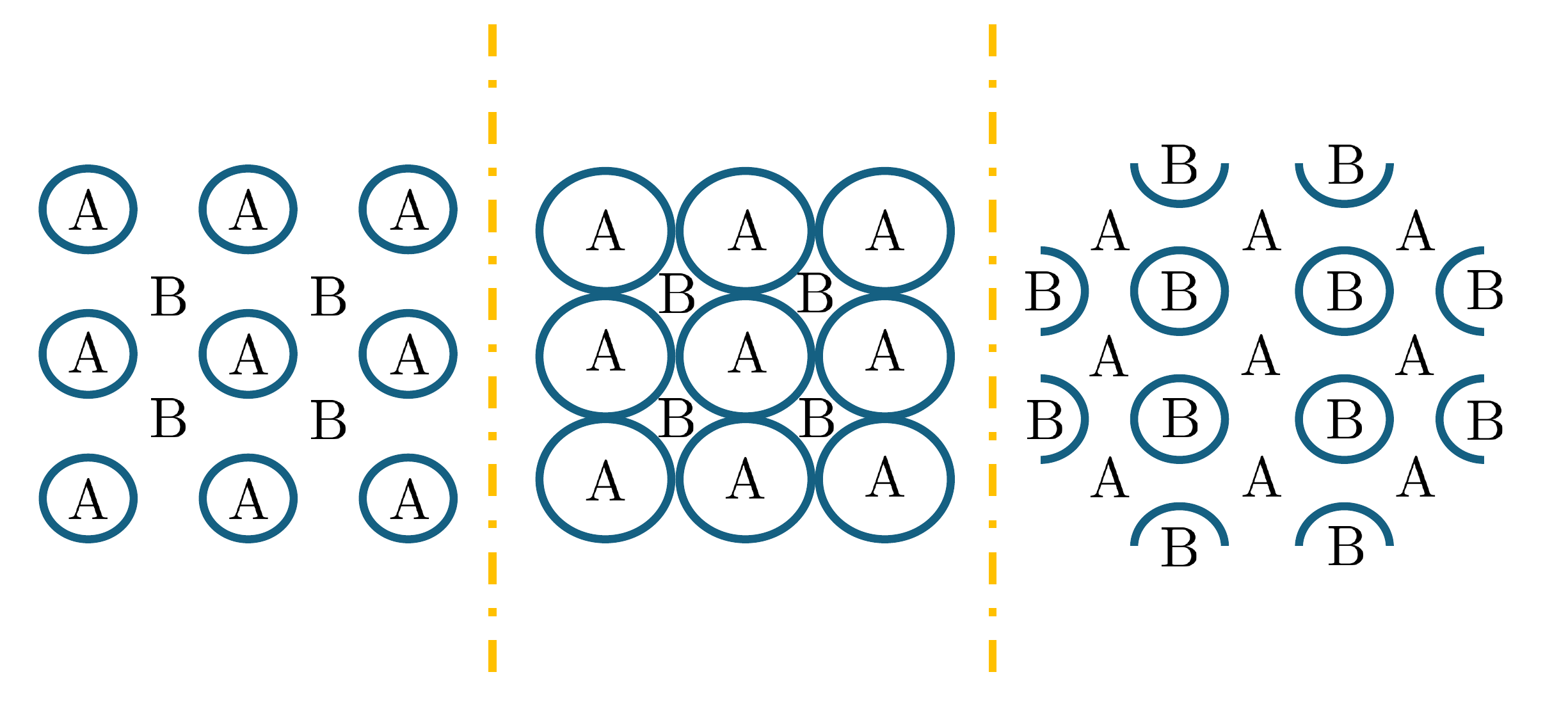}
\caption{Depiction of type II backgrounds in which a large number of GSO vortices are present. Depending on the number density of such vortices, this can lead to a bulk IIB theory with pockets of IIA (left), or conversely, a bulk IIA theory with pockets of IIB (right). This also suggests the possibility of a novel co-existence phase (middle) which would be interesting to investigate further.}
\label{fig:phasetransition}
\end{figure}

It is natural to also consider the statistical mechanics of GSO vortices, where energy minimization considerations need to be balanced against the tendency to increase entropy. This could potentially lead to a phase transition in which pockets of IIA and IIB simultaneously coexist, analogous to the Abrikosov vortex lattice that features in the transition out of the Meissner state in type II superconductors \cite{Abrikosov:1957wnz}. Or perhaps, even more speculatively, a full-fledged transition between type IIA and IIB driven by the proliferation of such GSO vortices. See Figure \ref{fig:phasetransition} for a depiction of this possibility.

More generally, one might hope to develop a unified theory of the domain walls and transitions between string vacua whose worldsheets are related by an orbifold. Such a picture might shed light on the target-space meaning of the orbifold procedure, which is not entirely clear in general (see \cite{Giaccari:2022xgs, Heckman:2024obe, Gomis:2025gzb} for more discussion).

In this direction, one further pair of string theories related by a worldsheet orbifold are the two supersymmetric heterotic string theories, with gauge algebras $\mathfrak{so}(32)$ and $\mathfrak{e}_8 \times e_8$. Indeed, the Het$_{\mathfrak{e}_8 \times \mathfrak{e_8}}$ string theory is obtained by performing one additional GSO projection relative to the Het$_{\mathfrak{so}(32)}$ string which acts separately on $\lambda^{A=1,...,16}$ and $\lambda^{A=17,..., 32}$. It would be very interesting if considerations similar to those in this paper could be used to study a potential domain wall between Het$_{\mathfrak{e}_8 \times \mathfrak{e_8}}$ and Het$_{\mathfrak{so}(32)}$.

We have presented further evidence that the $\mathrm{R}7$-branes support continuous higher-form potentials. In tandem with the likely appearance of strong coupling dynamics on their worldvolume, this provides some additional hints that the $\mathrm{R}7$-brane could potentially support a non-supersymmetric 8D conformal field theory (first conjectured in \cite{Dierigl:2022reg}), particularly due to their stability. It would clearly be exciting to explore this possibility further.\footnote{Another hint that the $\mathrm{R}7$-brane carries nontrivial gapless dynamics is that the merger of an $\Omega$ $\mathrm{R}7$-brane and an $F_L$ $\mathrm{R}7$-brane results in a supersymmetric 7-brane with $\mathfrak{so}(8)$ gauge theory \cite{Dierigl:2022reg}. While this 8D gauge theory is of course IR free, the fact that this merging can take place at all also supports the possibility of nontrivial dynamics on the $\mathrm{R}7$-brane worldvolume.}

Lastly, it is important to stress once again that our considerations in this paper point to the existence of a high-dimension, non-supersymmetric, and \textit{stable} 7-brane. This is likely to have many applications in the construction of non-supersymmetric string vacua, which paves the way for the development of a new class of string-phenomenological models. 

\section*{Acknowledgments}

JJH and ET thank M. Cveti\v{c} and G. Zoccorato for initial collaboration on attempts to build IIA/IIB walls in string theory and supergravity. JM and JPM thank H. Ooguri, B. Rayhaun, S.-H. Shao, and A. Sharon for initial collaboration on related work on vortices from the perspective of the worldsheet. We thank M. Cveti\v{c}, A. Debray, M. Delgado, M. Dierigl, M. Hopkins, J. Kaidi, F. Mignosa, M. Montero, H. Ooguri, A. Sen, A. Uranga, C. Vafa, X. Yin, X. Yu, and G. Zoccarato for helpful discussions. JJH and JM thank the 2023 meeting of the Simons Collaboration on Global Categorical Symmetries for hospitality during part of this work. JM, JPM, and ET thank the Aspen Center for Physics, which is supported by National Science Foundation grant PHY-2210452, for their hospitality during part of this work. The work of JJH is supported by DOE (HEP) Award DE-SC0013528 as well as by BSF grant 2022100. The work of JJH is also supported in part by a University Research Foundation grant at the University of Pennsylvania. JM is supported by DOE (HEP) Award DE-SC0011632. The work of ET is supported in part by the ERC Starting Grant QGuide-101042568 - StG 2021.

\appendix

\section{Backfiring Bosonization and the GSO Projection}\label{app:BACKFIRING}

In this appendix we comment on a technical subtlety in the GSO projection arising from a potential gravitational anomaly in performing the sum over spin structures.\footnote{We thank M. Montero and X. Yin for discussions on this subtlety.} As explained in \cite{BoyleSmith:2024qgx}, for a general 2D fermionic CFT, there is a subtle distinction between gauging the fermion parity symmetry $(-1)^f$ and summing over spin structures. The anomaly in summing over spin structures is captured by a mod-16 anomaly, while the anomaly in gauging fermion parity as a $\mathbb{Z}_2$ symmetry is merely defined mod-8.\footnote{These anomalies arise from the Anderson duals of bordism groups. For summing over spin structures, the relevant bordism calculation is the quotient group $\Omega_4^{\rm SO}/\Omega_4^{\rm Spin} = \mathbb{Z}_{16}$, while the relevant group for gauging fermion parity is $\Omega_3^{\rm Spin}(B \mathbb{Z}_2) = \mathbb{Z}_8$.} The contribution to either anomaly of any fermionic degree of freedom is given by $2 c_- = 2 (c_L - c_R)$, and the associated anomaly for summing over left-moving spin structures/gauging $(-1)^{f_L}$ is simply $2c_L$. When the total anomaly is 0 mod 16, there is no distinction between gauging fermion parity and summing over spin structures, and the end result is a bosonic CFT. However, when the anomaly is 8 mod 16, one may only gauge fermion parity, resulting in a new fermionic CFT. The key point is that when the anomaly is 8 mod 16, the Ramond sector states of the original theory have half-integer worldsheet spin, so that the resulting theory still requires a spin structure in order to be unambiguously defined on a general Riemann surface. One may say that, when the anomaly is 8 mod 16, bosonization has ``backfired.''

What role does this effect play on the string worldsheet, say, for the Type II string? Surprisingly, in the RNS formalism, the answer would at first glance appear to depend on the choice of gauge. In the covariant formalism, one has 10 chiral fermions $\psi^I$, each with $c_L = 1/2$, for a total matter anomaly $2 c_L = 10$. However, one must also include a contribution from the $\beta\gamma$ ghost system, which carry worldsheet fermion number. The ghost contribution to the anomaly is $2 c_L = 22$, for a total anomaly of $2 c_L = 32 \equiv 0$ mod 16, and so bosonization does not backfire in the covariant RNS formalism. This is compatible with the requirement that vertex operators, arising from operators of conformal weight $(1,1)$, be worldsheet bosons.

The situation seems different in lightcone gauge, where one simply has a matter theory consisting of 8 transverse chiral fermions $\psi^i$ and no ghost sector. Thus, in lightcone gauge, the total anomaly is $2 c_L = 8$, and so bosonization does indeed backfire. In fact, as described in \cite[Section 2.2]{BoyleSmith:2024qgx} (see also \cite{Tong:2019bbk}), the result of gauging fermion parity in lightcone gauge is again described by 8 chiral worldsheet fermions, now transforming in the spinor (or conjugate-spinor) representation of the transverse ${\rm Spin}(8)$ rotation group.\footnote{The chirality of the spinor representation is exchanged upon stacking with the 2D Arf TQFT.} This matter content precisely matches the lightcone-gauge Green-Schwarz string, whose torus partition function is indeed sensitive to the choice of spin structure, with the relevant choice being given by choosing periodic boundary conditions around both cycles of the torus.

How can the answer to whether bosonization backfires depend on the choice of worldsheet gauge? Moreover, there seems to be an issue with the conclusion that bosonization backfires in lightcone gauge: all local operators in the Ramond sector have half-integer worldsheet spin, and so one cannot construct vertex operators.

In fact, both issues are resolved by realizing that lightcone gauge is, globally, a singular gauge choice. In radial quantization, lightcone gauge may not be extended over the origin, and so one should not seek to define vertex operators in lightcone gauge. There is no issue with level matching on the cylinder, as the Casmir energy ensures that states still have integer eigenvalues under rotation of the cylinder. There is no tension with the covariant formalism, as the two gauges are not globally compatible.

A further subtlety is that, even in the covariant RNS string, the GSO projection does not actually arise from either gauging fermion parity or from summing over spin structures. Instead, one must integrate over the moduli spaces of super Riemann surfaces (see e.g., \cite{Witten:2012bh}), which may not have a holomorphic projection onto the moduli space of Riemann surfaces with a given genus and spin structure \cite{Donagi:2013dua}. As a result, one may not first perform the GSO projection and only then integrate over moduli. Instead, as explained in \cite{Witten:2012bh}, the GSO projection arises in the degeneration limit, where the sum over superconformal structures reduces to a sum over spin structures on the degenerating cycle. Nevertheless, the worldsheet supergravity theory still admits a global $\mathbb{Z}_2$ symmetry that assigns charge to the Ramond sector, corresponding to the quantum symmetry of the GSO projection. Moreover, each connected component of the moduli space of super Riemann surfaces corresponds to either an even or odd spin structure, as can be determined by setting the odd moduli to zero (see for instance section 2.1 of \cite{Witten:2012ga}), so one can still unambiguously define the $\mathrm{Arf}$ invariant for each super Riemann surface.

\section{Extension of the Type I Gauge Group}\label{app:EXTENSION}

In this appendix we describe in greater detail how the Type I gauge group is extended from $\mathrm{SO}(32)/\mathbb{Z}_2$ to $\mathrm{Spin}(32)/\mathbb{Z}_2$. In Section \ref{sec:Type_I_review} we described this extension as the result of the $\mathbb{Z}_2$ remnant of the RR field $C_8$, which gauges a magnetic $\mathbb{Z}_2$ 7-form symmetry of the $\mathrm{SO}(32)/\mathbb{Z}_2$ gauge theory. From this description, it is not immediately clear how to describe the extension in a duality frame involving only the RR fields $C_0, C_2, C_4$, or in a democratic formalism where all RR fields are included with an overall self-duality constraint. One puzzle is that the electromagnetic dual of a $\mathbb{Z}_2$ 8-form gauge field in 10 dimensions is a $\mathbb{Z}_2$ 1-form gauge field, which is precisely the $\mathbb{Z}_2$ gauge field that extends $\mathrm{SO}(32)/\mathbb{Z}_2$ to $\mathrm{Spin}(32)/\mathbb{Z}_2$. However, no such field seems to arise from the RR fields $C_0$ or $C_2$, which transform as
\begin{equation}\label{eq:Omega_on_Cs}
    C_0 \mapsto - C_0, \quad C_2 \mapsto C_2,
\end{equation}
under $\Omega$, leaving behind a $\mathbb{Z}_2$ valued parameter $C_0 = 0, \pi$, and a full $U(1)$ 2-form gauge field $C_2$.

The resolution is that, as explained in \cite{Gukov:1999yn}, the orientifold procedure involves taking equivariant cohomology, or equivalently, homotopy fixed points of the $\mathbb{Z}_2$ action of $\Omega$ on Type IIB field configurations. Thus, it is not enough that $C_2$ happens to equal its image under $\Omega$, we must instead choose a gauge parameter $\Lambda_1$ witnessing a gauge equivalence
\begin{equation}\label{eq:witness_eq}
    \Omega(C_2) = C_2 \sim C_2 + d \Lambda_1,
\end{equation}
which moreover satisfies $2 \Lambda_1 = d \lambda_0$ for some $\lambda_0$ valued in $\mathbb{R}/2 \pi \mathbb{Z}$ (as we are looking for $\mathbb{Z}_2$ homotopy fixed points, not $\mathbb{Z}$ homotopy fixed points). Thus, the $\Omega$-equivariance data for the RR field $C_2$ includes the $\mathbb{Z}_2$ 1-form gauge field $\Lambda_1$.

We claim that the gauge field $\Lambda_1$ is the electromagnetic dual of the $\mathbb{Z}_2$ 8-form gauge field arising from $C_8$, and is thus the 1-form $\mathbb{Z}_2$ gauge field that extends $\mathrm{SO}(32)/\mathbb{Z}_2$ to $\mathrm{Spin}(32)/\mathbb{Z}_2$. In order to see this, we use a democratic formalism in which all RR fields are included as a differential $K$-theory gauge field for the shift $\Sigma^{-1} KU$ of the complex $K$-theory spectrum. The homotopy groups $\pi_{\rm odd} (\Sigma^{-1} KU) = \mathbb{Z}$ correspond to the RR field strengths $F_{\rm odd}$ of Type IIB, and the self-duality constraint is imposed via the Anderson self-duality\footnote{Here, $I_{\mathbb{Z}}X$ refers to the Anderson dual of a spectrum $X$ \cite{Lurie:2012, beaudry2018guidecomputingstablehomotopy}.}
\begin{equation}\label{eq:IIB_Anderson}
    \Sigma^{10} I_\mathbb{Z} \Sigma^{-1} KU =\Sigma^{11} I_\mathbb{Z} KU = \Sigma^{11} KU = \Sigma^{-1} KU,
\end{equation}
using the Anderson self-duality of $KU$ together with Bott periodicity.

Now, the orientifold projection involves taking homotopy fixed points for the $\mathbb{Z}_2$ action of $\Omega$ on $\Sigma^{-1} KU$, which acts via (the shift of) complex conjugation as in \eqref{eq:Omega_on_Cs}. The homotopy fixed point spectrum has been computed (see, e.g., \cite{Heard_2014}), and is given by $\Sigma^{-1} KO$, the $K$-theory spectrum relevant for Type I. Moreover, as shown in \cite{Heard_2014}, the Anderson self-duality $I_{\mathbb{Z}} KU = \Sigma^4 KU$ holds $\mathbb{Z}_2$-equivariantly (which implies $I_{\mathbb{Z}} KO = \Sigma^4 KO$), and so we have the Anderson self-duality
\begin{equation}\label{eq:I_Anderson}
    \Sigma^{10} I_\mathbb{Z} \Sigma^{-1} KO = \Sigma^{11} I_\mathbb{Z} KO = \Sigma^{15} KO = \Sigma^{-1} KO,
\end{equation}
using $I_{\mathbb{Z}} KO = \Sigma^4 KO$ and Bott periodicity. Equation \eqref{eq:I_Anderson} is the appropriate self-duality constraint for the $\Omega$-projected RR fields in Type I.

To connect with our discussion of the gauge fields $C_8$ and $\Lambda_1$, we simply identify them as the gauge fields corresponding to the homotopy groups
\begin{equation}
    \pi_1 (\Sigma^{-1} KO) = \mathbb{Z}_2, \quad \pi_8 (\Sigma^{-1} KO) = \mathbb{Z}_2.
\end{equation}
This follows from the homotopy fixed point spectral sequence\footnote{\label{footnote:DrbrayNotes}See also \cite{DebrayNotes} for a helpful account of the homotopy fixed point spectral sequence.}
\begin{equation}
    H^p_{\rm grp}(\mathbb{Z}_2; \pi_q(\Sigma^{-1} KU)) \Rightarrow \pi_{q - p}(\Sigma^{-1} KO),
\end{equation}
as computed in \cite{Heard_2014}. Indeed, $\pi_8(\Sigma^{-1} KO) = \mathbb{Z}_2$ arises from the cohomology group
\begin{equation}
    H^1_{\rm grp}(\mathbb{Z}_2; \pi_9(\Sigma^{-1} KU)) = \mathbb{Z}_2,
\end{equation}
corresponding to taking $\mathbb{Z}_2$ torsion $C_8$, while $\pi_1(\Sigma^{-1} KO) = \mathbb{Z}_2$ arises from
\begin{equation}
    H^2_{\rm grp}(\mathbb{Z}_2; \pi_3(\Sigma^{-1} KU)) = \mathbb{Z}_2,
\end{equation}
corresponding to the gauge parameter $\Lambda_1$ witnessing the equivalence $\Omega(C_2) \sim C_2$ as in \eqref{eq:witness_eq}. This calculation should be compared to the equivariant cohomology groups appearing in the calculation of discrete torsion for orientifold planes \cite{Witten:1998cd, Bergman:2001rp}, except that the result is the discrete gauge field $\Lambda_1$ rather than a discrete parameter.

\section{Relation Between RR Fields in Type IIA and IIB} \label{app:MOD2}

In this appendix we establish that, under the orbifold by $(-1)^{F_L}$, the discrete remnants of the RR potentials from IIA (resp. IIB) are captured by the mod-2 reduction of the fluxes of IIB (resp. IIA), i.e., that we have
\begin{equation}\label{eq:mod2relation_copied}
C_{2k-1} = \frac{1}{2} F_{2k-1} \,\,\, \mathrm{mod} \, 2\pi  \,\,\, \text{(IIB)} \\\,\,\, \text{and} \,\,\, C_{2k} = \frac{1}{2}F_{2k}\,\,\, \mathrm{mod} \, 2\pi \,\,\, \text{(IIA)}, 
\end{equation} 
as in line \eqref{eq:mod2relation}. In order to show this, we use that the RR field configurations of IIB (resp. IIA) are given by the $\mathbb{Z}_2$-equivariant RR field configurations of IIA (resp. IIB) with respect to the action of $(-1)^{F_L}$, as described as differential cocycles for $KU$ (resp. $\Sigma KU$). We will focus on the first equation in \eqref{eq:mod2relation_copied}, relating the remnants of RR potentials from IIA to the RR fluxes in IIB; the other follows from entirely analogous arguments.

Thus, let us consider the action of $(-1)^{F_L}$ on $KU$ by negation, i.e., mapping a virtual vector bundle $(V, W) \mapsto -(V, W) = (W, V)$.\footnote{The $K$-theory class of the pair $(V, W)$ corresponds to the formal difference $[V] - [W]$.} Crucially, $(-1)^{F_L}$ acts by a sign on all homotopy groups of $KU$, in accordance with its action on the RR fields. This $\mathbb{Z}_2$ action was studied in \cite{Atiyah:2003qq}, who showed that the associated $\mathbb{Z}_2$-equivariant theory $KU_\pm$ associated to the $\mathbb{Z}_2$ action of $(-1)^{F_L}$ on $KU$ reduces to the shift $\Sigma KU$ when the action of $(-1)^{F_L}$ on spacetime is trivial.\footnote{We thank A. Debray and M. Hopkins for discussions on this point.} To build intuition for this result, note that a virtual vector bundle $(V, W)$ equipped with equivariance data $\lambda : (V, W) \to (W, V), \lambda^2 = 1$ is the same data as a module for the complex Clifford algebra $Cl_1$, which provides a model for the shift $\Sigma KU$ (see \cite{AtiyahBottShapiro}). In worldsheet terms the appearance of this algebra is also connected with the appearance of the Majorana mode localized at the  GSO interface between IIA and IIB discussed in Section \ref{sec:LONGSTRING}.

\begin{figure}
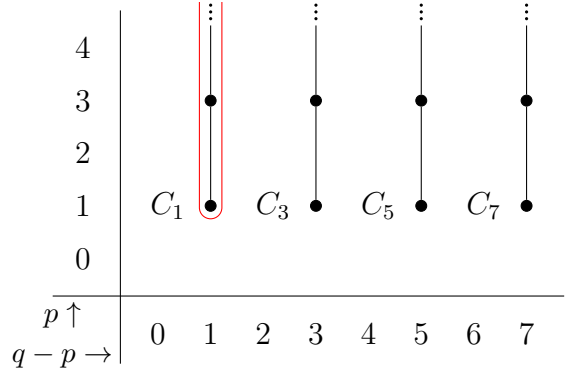

\centering
\begin{sseqdata}[name=hfpss, classes={fill, show name = left}, xrange={0}{7}, yrange={0}{4}, scale=0.7, Adams grading, >=stealth,
x label = {$\displaystyle{p\uparrow \atop q-p\rightarrow}$},
x label style = {font = \small, xshift = -20.5ex, yshift=3ex}]
\class[name=C_1](1, 1)
\class(1, 3)\structline
\class(1, 5)\structline
\class[name=C_3](3, 1)
\class(3, 3)\structline
\class(3, 5)\structline
\class[name=C_5](5, 1)
\class(5, 3)\structline
\class(5, 5)\structline
\class[name=C_7](7, 1)
\class(7, 3)\structline
\class(7, 5)\structline
\circleclasses[rounded rectangle,color=red](1,1)(1,5)
\end{sseqdata}
\printpage[name=hfpss, page=2]
\caption{The $E_2$ (and $E_\infty$) page of the homotopy fixed point spectral sequence for the $\mathbb{Z}_2$ action of $(-1)^{F_L}$ on $KU$. Each black dot represents the group $\mathbb{Z}_2$. As indicated, the $\mathbb{Z}_2$ classes are connected by extensions. The bottom node in each tower corresponds to the $\mathbb{Z}_2$ remnant of the corresponding IIA potential $C_{2k-1}$. Each tower corresponds to a IIB flux $F_{2k-1}$; for instance the tower circled in red corresponds to the flux $F_1$.}
\label{fig:FL_hfpss}
\end{figure}

In order to derive \eqref{eq:mod2relation_copied} from the result of \cite{Atiyah:2003qq} that $K_\pm$ reduces to $\Sigma KU$ when the $\mathbb{Z}_2$ action on spacetime is trivial, consider the homotopy fixed point spectral sequence
\begin{equation}
    E_{2}^{p,q} = H^{-p}_{\mathrm{grp}}(\mathbb{Z}_2; \pi_q(KU)) \Rightarrow \pi_{p+q}(KU^{\mathbb{Z}_2}),
\end{equation}
associated to the $\mathbb{Z}_2$ action of $(-1)^{F_L}$.\footnote{As in Footnote \ref{footnote:DrbrayNotes}, see \cite{DebrayNotes} for a helpful account of the homotopy fixed point spectral sequence.} While the homotopy fixed points of $(-1)^{F_L}$ acting on $KU$ are not quite the same thing as $\mathbb{Z}_2$-equivariant $K_\pm$ classes, they are closely related: the homotopy fixed points are a completion of the equivariant spectrum, as in the Atiyah-Segal completion theorem \cite{atiyah1969equivariant}. As $(-1)^{F_L}$ acts by a sign on all homotopy groups (since it acts by a sign on all RR fields), we have that the $E_2$ page is given by
\begin{equation}
    H^p_{\rm grp}(\mathbb{Z}_2; \pi_q( KU)) = \begin{cases} \mathbb{Z}_2, & p > 0 \text{ odd}, q \text{ even}, \\ 0, & \text{otherwise}\end{cases}
\end{equation}
where we emphasize that the coefficients, $\pi_q(KU)$, are non-trivial $\mathbb{Z}_2$-modules. See Figure \ref{fig:FL_hfpss} for a depiction of this spectral sequence. On degree reasons, there can be no differentials, and so the spectral sequence collapses at the $E_2$ page. We see that the (2-completions of the) homotopy groups $\pi_{\rm odd}(\Sigma KU) = \mathbb{Z}$ arise from extensions in the towers of $\mathbb{Z}_2$ classes on the $E_\infty$ page.

Physically, each class in $H^1_{\rm grp}(\mathbb{Z}_2; \pi_{2k}(KU)) = \mathbb{Z}_2$ arises from the $\mathbb{Z}_2$ remnant of the IIA potential $C_{2k-1}$, as indicated in Figure \ref{fig:FL_hfpss}. Moreover, the tower in $q-p=2k-1$ corresponding to $\pi_{2k-1}(\Sigma KU) = \mathbb{Z}$ corresponds to the IIB flux $F_{2k-1}$. As the class at the bottom of a tower of extensions corresponds to the mod-2 parity in the group $\mathbb{Z}$, we learn that the $\mathbb{Z}_2$ remnants of the RR potentials $C_{2k-1}$ can be identified with the mod-2 parity of the integer-quantized RR fluxes $F_{2k-1}$, as desired.

The lift of the RR fluxes $F_{2k-1}$ from $\mathbb{Z}_2$ to $\mathbb{Z}$ arises as a consequence of the cohomology groups $H^{2\ell+1}_{\rm grp}(\mathbb{Z}_2; \pi_{2(k + \ell)}(KU))$, for $\ell > 0$. As in the case of the field $\Lambda_1$ discussed in Appendix \ref{app:EXTENSION}, these groups correspond to equivariance data for the higher-form potentials $C_{2(k+\ell)-1}$. Interestingly, in order to lift all the way to $\mathbb{Z}$, we would need to discuss equivariance data for RR potentials $C_{\rm odd}$ of form degrees greater than the dimension of spacetime, which seem to lack any obvious physical interpretation.

\section{Lagrange Multipliers and Fluxes} \label{app:LAGMULT}

In this appendix we review the formulation of type II theories in terms of a path integral over fluxes and suitable Lagrange multipliers. See e.g., references \cite{Witten:1995gf, Bergshoeff:2001pv, Kapustin:2010} for some additional discussion of this formulation. In what follows we suppress all contributions to the (pseudo-)Lagrangians other than those directly necessary to discuss the RR sector of the theory.

Recall that IIA\ contains the RR potentials $C_{1}$ and $C_{3}$,
with corresponding field strengths $F_{2}$ and $F_{4}$. We can also include a
zero-form flux $F_{0}$ associated with the Romans mass. We present this data
by including Lagrange multiplier fields $A_{5}$ and $A_{7}$ and a source
$\Lambda_{10}$ (for the Romans mass)\ so that the Lagrangian density takes the
form (see e.g., \cite{Witten:1995gf, Bergshoeff:2001pv, Kapustin:2010}):
\begin{align}
L_{\text{IIA}} &  = \frac{1}{2}F_{0}\wedge\ast F_{0}+ \frac{1}{2}F_{2}\wedge\ast F_{2}+\frac{1}{2}F_{4}\wedge\ast
F_{4}\\
&  -F_{0}\wedge\Lambda_{10}-F_{2}\wedge dA_{7}-F_{4}\wedge dA_{5}.
\end{align}
In this formulation the path integral is over the $F_{k}$ as well as the
Lagrange multipliers $A_{m}$. Here, the ``fluxes'' are valued in $\mathbb{R}$ and the Lagrange multipliers are valued in $U(1)$. Flux quantization is only recovered once we impose the equations of motion. 
Indeed, the local formulation in terms of gauge potentials follows from the equations of motion: 
the equations of motion for $F_{2}$ and $F_{4}$ tell us that we have local presentations of the
form $\ast F_{2}=dA_{7}$ and $\ast F_{4}=dA_{5}$, while the equation of motion
for $A_{7}$ and $A_{5}$ enforce $dF_{2}=0$ and $dF_{4}=0$. Said differently, on-shell 
$A_{7}$ and $A_{5}$ are simply the RR\ potentials for the 8-form and 6-form
fluxes. The equation of motion for $F_{0}$ simply fixes the value of a
spacetime filling flux, as specified by $\Lambda_{10}$.

We can also reverse the roles, working in terms of the Hodge dual fluxes:
\begin{align}
L_{\text{IIA}}^{\prime} &  =\frac{1}{2}F_{10}\wedge\ast F_{10}+\frac{1}{2}F_{8}\wedge\ast
F_{8}+\frac{1}{2}F_{6}\wedge\ast F_{6}\\
&  -F_{10}\wedge\Lambda_{0}-F_{8}\wedge dA_{1}-F_{6}\wedge dA_{3},
\end{align}
and the path integral is over the $F_{k}$ as well as the Lagrange multipliers
$A_{m}$. Observe that the equations of motion for $F_{8}$ and $F_{6}$ tell us
that we have local presentations of the form $\ast F_{8}=dA_{1}$ and $\ast
F_{6}=dA_{3}$, while the equations of motion for $A_{1}$ and $A_{3}$ enforce
$dF_{8}=0$ and $dF_{6}=0$. Said differently, on-shell $A_{1}$ and $A_{3}$ are simply
the RR potentials for the 2-form and 4-form fluxes. The equation of motion for
$F_{10}$ fixes the value of the cosmological constant.

Starting in IIB, we can likewise introduce the pseudo-Lagrangian:\footnote{It
is a pseudo-Lagrangian in the sense that we impose the self-duality condition
for the $5$-form flux after obtaining the equations of motion. For further
discussion see in particular section 9 of reference \cite{Belov:2006xj}.}
\begin{align}
L_{\text{IIB}} &  =\frac{1}{2}F_{1}\wedge\ast F_{1}+\frac{1}{2}F_{3}\wedge\ast F_{3}+\frac{1}{2}F_{5}\wedge\ast
F_{5}\\
&  -F_{1}\wedge dA_{8}-F_{3}\wedge dA_{6}-F_{5}\wedge dA_{4},
\end{align}
from which we recover the analogous statements for IIB\ fluxes and
RR\ potentials, and where the condition $\ast F_{5}=F_{5}$ is imposed after
obtaining the equations of motion. The Hodge dual formulation is given by:%
\begin{align}
L_{\text{IIB}}^{\prime} &  =\frac{1}{2}F_{9}\wedge\ast F_{9}+\frac{1}{2}F_{7}\wedge\ast F_{7}%
+\frac{1}{2}F_{5}\wedge\ast F_{5}\\
&  -F_{9}\wedge dA_{0}-F_{7}\wedge dA_{2}-F_{5}\wedge dA_{4},
\end{align}
in the obvious notation. Observe that in both the \textquotedblleft
electric\textquotedblright\ and \textquotedblleft magnetic\textquotedblright%
\ pseudo-Lagrangians, the Lagrange multiplier $A_{4}$ appears because the
five-form field strength is self-dual on-shell. In this sense there is no
canonical electric/magnetic split, a point emphasized in \cite{Belov:2006xj}.

\section{Fluxbranes and Non-BPS Branes} \label{app:FLUXBRANE}

In this appendix we discuss the relation between non-BPS branes, brane/anti-brane configurations, and fluxbranes \cite{Gutperle:2001mb, Emparan:2001gm}.\footnote{Fluxbranes provide a way to engineer the topological symmetry operators for continuous symmetries \cite{Cvetic:2023plv}. Brane/anti-brane configurations can serve a similar purpose, as noted in Appendix A of reference \cite{Cvetic:2023plv}, as can non-BPS branes \cite{Bergman:2024aly, Calvo:2025kjh}. As we argue in this Appendix, these different configurations are all identical.}

Recall that a fluxbrane is merely a higher-/lower-dimensional generalization of a fluxtube, just as D-branes are higher-/lower-dimensional generalizations of charged particles \cite{Gutperle:2001mb, Emparan:2001gm, Cvetic:2023plv}. In particular, an RR flux-$p$-brane (or $p$-fluxbrane) configuration is characterized by 
\begin{equation}\label{eq:flux_delta}
    *F_{p+1}=\eta \delta_{\Sigma_{p+1}} 
\end{equation}
where $\Sigma_{p+1}$ is the flux-$p$-brane worldvolume. As pure gauge field configurations, the parameter $\eta$ is well-defined as a real number. However, one can annihilate fluxbranes configurations with $\eta=2\pi n$ by having one end of the flux-$p$-brane terminate on $n$ $\mathrm{D}(p-1)$-branes and the other end terminate on $n$ $\overline{D(p-1)}$-branes. Thus, $\eta$ only defines a conserved $p$-brane charge modulo $2\pi$,\footnote{One can view $\eta$ mod $2 \pi$ as the gauge charge associated to the gauging of the global $p$-form $\mathbb{Z}$ symmetry we would have had if $C_{p}$ were an $\mathbb{R}$ gauge field, rather than a ${\rm U}(1)$ gauge field.} which can equivalently be defined by
\begin{equation}
    \eta = \int_{S^{8-p}} C_{8 - p},
\end{equation}
where $S^{8-p}$ is the sphere linking the brane worldvolume, and where $d C_{8 - p} = \ast F_{p + 1}$ is the electromagnetically dual RR potential. Of course, one can also consider NSNS fluxbranes, which terminate on F$1$-strings or $\mathrm{NS}5$-branes when $\eta=2\pi n$, as well as M-theory fluxbranes.

In general, in a Coulomb phase where the gauge fields are massless, delta-function localized fluxbranes as in \eqref{eq:flux_delta} are off-shell (and non-BPS) configurations which will dynamically spread in the transverse directions. The result of this spreading process is a dilute gauge field configuration with total flux $\eta$, which we will refer to as a dilute flux-$p$-brane. By contrast, in a confining phase (or, dually, a Higgs phase for the dual field), fluxbranes are genuine, localized branes with finite tensions. For example, in a superconductor, magnetic fluxbranes are simply magnetic flux tubes, with a finite string tension.

\begin{figure}[t!]
\centering
\includegraphics[scale=0.6, trim={0 6cm 0 6cm}]{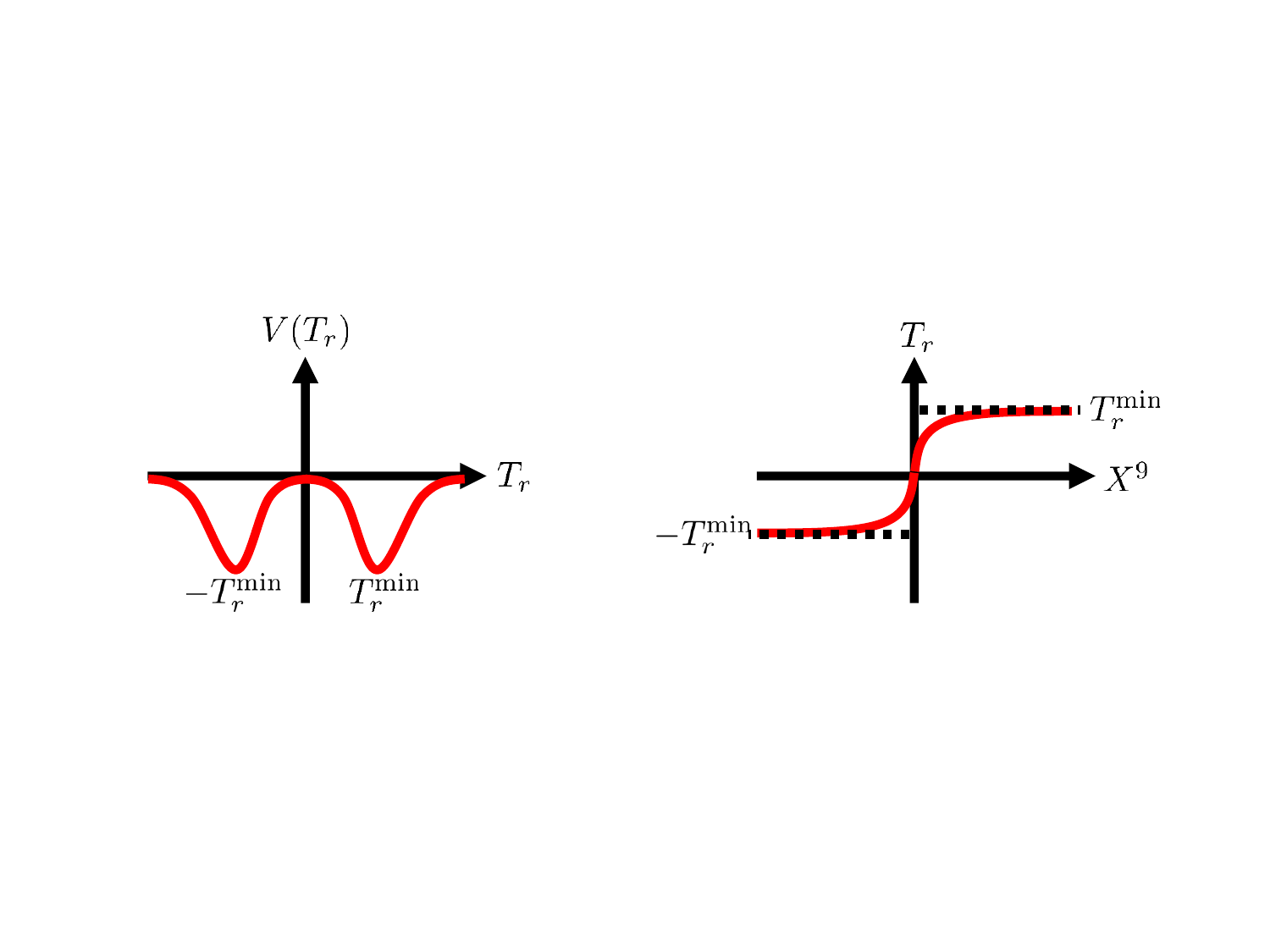}
\caption{Depiction of the real tachyon profile $T_r$ for a non-BPS D-brane. On the left we show the effective potential for the tachyon, including its minima at $\pm T_{r}^{\mathrm{min}}$. On the right we show the kink profile of the tachyon as it interpolates from $-T_{r}^{\mathrm{min}}$ to $T_{r}^{\mathrm{min}}$. The potential for the complex tachyon of a brane/anti-brane configuration can be obtained in a similar fashion by a surface of revolution.}
\label{fig:TachyonPicture}
\end{figure}

The main purpose of this appendix is to argue that the non-BPS D$p$-branes of Type II string theory are precisely the same objects as RR flux-$p$-branes. Additionally, their fluxbrane charge $\eta$ is determined by the value of the worldvolume real tachyon field $T_r$, in such a way that $\eta$ goes from 0 to $2 \pi$ as the tachyon $T_r$ goes from $-T^{\mathrm{min}}_r$ to $+T^{\mathrm{min}}_r$, where $T^{\mathrm{min}}_r$ is the minimum of the tachyon potential (see Figure \ref{fig:TachyonPicture}). Moreover, the top of the tachyon potential $T_r = 0$ corresponds to a fluxbrane charge $\eta = \pi$, as was previously argued in \cite{Harvey:2000qu}. If one starts with a non-BPS D$p$-brane with some initial value $T_r = T_0$, the end result of tachyon decay is a dilute fluxbrane with fluxbrane charge $\eta$ determined by $T_0$. Many of these statements are essentially contained in \cite{Bergman:2024aly}, however, the interpretation given there is different. As an aside, let us note that the identification of the non-BPS D-branes with RR flux-$p$-branes is consistent with the conjecture of \cite[Section 7.2]{Intriligator:2000pk} that the potential for $T_r$ is actually periodic with an infinite sequence of minima, which would correspond to $\eta = 2 \pi n$ for $n \in \mathbb{Z}$.

\begin{figure}[t!]
\centering
\includegraphics[scale=0.4, trim={0 0cm 0 0cm}]{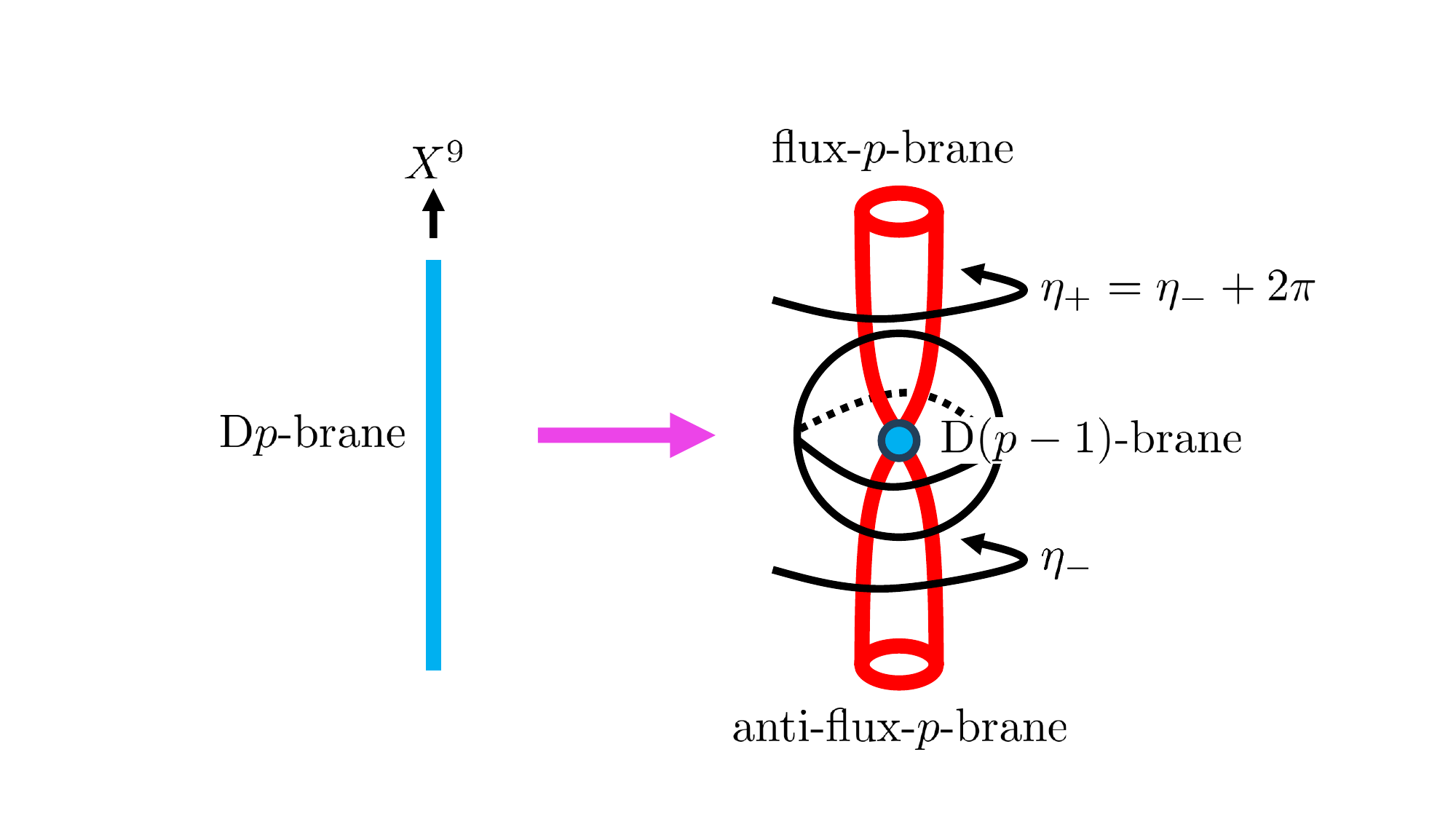}
\caption{By allowing the tachyon on a non-BPS D$p$-brane to decay, we can show that its fluxbrane charge must be given by $\eta = \pi$. On the left, a non-BPS D$p$-brane is extended in the spatial direction $X^9$. On the right, the real tachyon field $T_r$ is allowed to decay towards its positive minimum along the positive $X^9$-axis, and towards its negative minimum along the negative $X^9$-axis. The end result is a tachyon kink configuration, which produces a BPS D$(p-1)$-brane at $X^9 = 0$. The flux through a sphere linking the D$(p-1)$-brane must be $2 \pi$, which implies that the flux upwards through the northern and southern hemispheres must differ by $2 \pi$.}
\label{fig:NonBPSeta}
\end{figure}

One simple way to reach these conclusions relies only on symmetry. The non-BPS D$p$-brane is even under $(-1)^{F_L}$, and so its RR fluxbrane charge must be either $0$ or $\pi$ mod $2 \pi$, since the RR fields are odd under $(-1)^{F_L}$. Consider a non-BPS D$p$-brane extended along at least one spatial direction, say $X^9$, as depicted in Figure \ref{fig:NonBPSeta}. At first, we set the tachyon to zero, $T_r = 0$, along the entire D$p$-brane. If we then allow $T_r$ to dynamically decay towards its positive minimum $T^{\mathrm{min}}_r > 0$ along the positive $X^9$-axis, and towards its negative minimum $- T^{\mathrm{min}}_{r} < 0$ along the negative $X^9$-axis, we obtain a tachyon kink configuration that has been identified with a BPS D$(p-1)$-brane \cite{Sen:1999mg}. This D$(p-1)$-brane must source one unit of flux of $\ast F_{p+1}$, which implies that the fluxbrane charges $\eta_\pm$ of the resulting gauge field configurations along the positive and negative $X^9$-axis must differ by $2 \pi$, i.e., $\eta_+ = \eta_- + 2 \pi$. But, since the real tachyon $T_{r}$ is odd under $(-1)^{F_L}$, we must further have $\eta_- = - \eta_+$, which implies that $2 \eta_\pm = \pm 2 \pi$, or in other words, $\eta_\pm = \pm \pi$, as claimed.

It is important to stress the distinction between tuning the asymptotic value of $T_r$ as a parameter of the worldvolume theory of the non-BPS D$p$-brane\footnote{This parameter corresponds to the continuous parameter controlling the symmetry operators constructed by placing the non-BPS D-branes at infinity, according to the picture developed in \cite{Apruzzi:2022rei, GarciaEtxebarria:2022vzq, Heckman:2022muc, Heckman:2022xgu}. One might worry that the tachyon field $T_r$ has a nontrivial potential; however, in moving the brane to infinity one is taking a decoupling limit in which stringy dynamics have been stripped away. A continuous parameter space then emerges.} and allowing the tachyon field $T_r$ to dynamically decay. In the first case, we have a one-parameter family of D-brane configurations, with variable fluxbrane charge $\eta$. In the second, however, the total fluxbrane charge $\eta$ must be constant (mod $2 \pi$) throughout the decay; indeed, it is a conserved charge. While the value of the tachyon field $T_r$ relaxes to its minimum, closed-string fields are excited, in order to maintain the total energy of the configuration \cite{Sen:1999mg}. These closed-string fields must absorb the fluxbrane charge, producing a smooth gauge field configuration with increasing flux. Towards the end of the decay, the entire fluxbrane charge $\eta$---determined by the initial value $T_r = T_0$ of the tachyon field---is carried by the smooth gauge field configuration.\footnote{The ultimate fate of the non-BPS D-brane is thus a ever-diluting gauge flux, with constant total fluxbrane charge at any finite time.} The evolving value of the tachyon field $T_r$ corresponds to the (initially) delta-function localized contribution to the fluxbrane charge, not the asymptotic value measured in any particular field configuration.

We provide two further arguments that the non-BPS D-branes can be identified with fluxbranes, with $\eta = \pi$ when the tachyon field $T_r$ is set to zero\footnote{See section 6 of \cite{Harvey:2000qu} for an earlier argument.}. First, we note that a BPS D$(7-p)$-brane, which is magnetically charged under $C_p$, picks up a sign when carried around a non-BPS D$p$-brane. This can be seen by an argument analogous to the one given in \cite[Section 4.1]{Gukov:1999yn}, where it was shown that the non-BPS D$0$-brane of Type I picks up a sign when carried around the Type I D$7$-brane. The basic argument given there is that the Ramond sector ground states in the $0-7$ and $7-0$ sectors form a two-state system, which undergo level crossing precisely when the branes are brought together, and which are subject to a Berry phase of $\pi$ upon carrying the branes around each other. In our case, we do not have an anti-D$(7-p)$-brane, but we also do not have the orientifold projection $\Omega$. Thus, we have the same Ramond sector ground states, and can derive the same Berry phase. This sign implies that the presence of a non-BPS D$p$-brane induces a monodromy of $\pi$ in $C_{8-p}$ around its worldvolume, or in other words, that the fluxbrane charge of the non-BPS D$p$-brane is $\eta = \pi$ mod $2 \pi$.

As a second line of analysis we now discuss how the same fluxbrane 
charge arises from the brane/anti-brane system, as well as from the non-BPS D-brane itself. 
To set notation, we consider a brane anti/brane system on $(p+2)$-dimensional subspace $\Sigma_{p+2}$ and seek field configurations which are quasi-localized on a subspace $\Sigma
_{p+1}\subset\Sigma_{p+2}$ which we take to be given by $X_{\bot}=0$ with $X_{\bot}$ a local coordinate. Since we are assuming the induced brane (be it a fluxbrane or a non-BPS D-brane)\ is
quasi-localized in this direction, the behavior far away from the region
$X_{\bot}=0$ remains outside our bailiwick.

The WZ terms of the brane/anti-brane system can in principle be captured by an appropriate superconnection, see e.g., references \cite{Quillen:1985, Kennedy:1999nn,
Vafa:2001qf, Szabo:2001yd, Antonelli:2019pzg} for definitions and calculations. Our aim here will be a bit more modest, and so we will attempt to constrain the structure of candidate topological couplings using bottom up considerations. With that in mind, we keep the minimal field content associated with a $U(1)_+ \times U(1)_{-}$ gauge group and a complex tachyon $T_c$ with charge $(+1,-1)$. Consider then the WZ coupling of the gauge fields on the brane/anti-brane stack to the pullback of the bulk RR potential $C_p$:
\begin{equation}
S_{\text{WZ}}^{\text{brane/anti-brane}}=\int_{\Sigma_{p+2}}C_{p}\wedge\left(
f_{+}-f_{-} \right)  +...,\label{eq:WZddbar}
\end{equation}
where locally, $f_{+} = da_{+}$ and $f_{-} = da_{-}$ with $a_{\pm}$ the vector potential for the gauge group $U(1)_{\pm}$.

How should we include the effects of the tachyon profile in any candidate topological terms? To argue for this, we note that in general, the tachyon has a non-trivial background profile, and so the axial $U(1)_{\mathrm{axial}}$ with vector potential given by the combination $a = a_{+} - a_{-}$ will end up getting broken. We can explicitly include the Goldstone mode eaten during this breaking process by taking into account the phase of the complex tachyon $T_c = \vert T_c \vert \exp(i \theta)$. Observe that gauge transformations act as:
\begin{equation}
a\mapsto a+d\lambda\text{ \ \ and \ \ }\theta\mapsto\theta-\lambda.
\end{equation}
In particular, the combination $a + d \theta$ is inert under such gauge transformations.

Returning to equation (\ref{eq:WZddbar}), observe that integration by parts of the local presentation $f_{+} - f_{-} = da$ would result in a term of the form $F_{p+1} \wedge a$, with $F_{p+1} = dC_{p}$. Integrating over $a$ in the $X_{\bot}$ direction would then produce a term $\eta F_{p+1}$ of precisely the sort expected for a fluxbrane (with charge $\eta$). Observe, however, that while $F_{p+1} \wedge a$ is indeed invariant under small gauge transformations, under large gauge transformations the putative value of $\eta$ would appear to shift, by an arbitrary amount which depends on $\lambda$. Throughout this paper we have implicitly assumed that fluxbrane charge is globally well-defined, and have presented a self-consistent picture compatible with this. Said differently, we expect that fluxbrane charge can be consistently measured far away from an object. Our conclusion is that the term $F_{p+1} \wedge a$ cannot by itself be the full story.

The minimal resolution of these puzzles compatible with these considerations is that there is in fact another coupling, as required by gauge invariance:
\begin{equation}\label{eq:WZimproved}
S_{\text{WZ}}^{\text{brane/anti-brane}}=\int_{\Sigma_{p+2}}F_{p+1}%
\wedge\left(  a+d\theta\right)  +...,
\end{equation}
namely we have the covariant derivative of the tachyon phase. Integrating this over a direction transverse to the fluxbrane then results in an unambiguous assignment for $\eta$.

Before proceeding further, it is worth commenting that the presentation just given is singular near the core of a candidate tachyon solution, since, for example, in a kink profile for the tachyon $\vert T_c \vert \rightarrow 0$, the phase of the tachyon discontinuously jumps. Additionally, it is worth emphasizing that in this regime an effective field theory analysis is inappropriate anyway (one ought to instead analyze the entire process in string field theory).

That being said, it is worth noting that we can deform any candidate field configuration in the complex $T_c$-plane away from the origin, and our expectation is that topological couplings will be robust under such deformations. Additionally, observe that even in configurations such as $T_c \sim \tanh{ (X_\bot / \ell)}$ where the phase $\theta(X_\bot)$ discontinuously jumps, $\exp \left(i \pi \int_{\Sigma_{p+1}} F_{p+1} \right)$ smoothly connects to $\exp \left(-i \pi \int_{\Sigma_{p+1}} F_{p+1} \right)$.

In any case, we leave a full treatment of these effects for future work. For now, we simply observe that the proposed coupling of line (\ref{eq:WZimproved}) allows us to capture the fluxbrane charge of both a fluxbrane and a non-BPS $\mathrm{D}p$-brane. Indeed, by inspection, the standard fluxbrane WZ term (such as the ones considered in \cite{Cvetic:2023plv}) comes from integrating $a$ across the
$X_{\bot}$-direction in a tachyon profile where $\theta$ integrates to zero. 
Likewise, the standard non-BPS\ D-brane kink solution is
constructed purely from the real tachyon profile $T_{c}\sim\tanh(X_{\bot}%
/\ell)$ in a profile where we switch off $a$.\footnote{Observe that this choice is rather singular, but there is still a jump in the phase $\theta$ of previsely $\pi$.} Note also that the fluxbrane charge
of the non-BPS D-brane is $\pi$ mod $2\pi$, in accord with the analysis
presented earlier. One can also consider more general combinations which one might interpret as combinations of non-BPS D-branes dressed by fluxbranes which carry more general asymptotic values for $\eta$.

Consider now the case of a non-BPS D$p$-brane. This brane also has a ``WZ-like term'' in which one's ignorance of instabilities is packaged in terms of a real open-string tachyon $T_{r}$ (see \cite{Sen:2004nf}):
\begin{equation}
S_{\mathrm{WZ}}^{\mathrm{Non-BPS}} = \underset{\Sigma_{p+1}}{\int} W(T_r) \, C\wedge\exp f,
\end{equation}
in the obvious notation. Here, the precise function $W(T_r)$ is (as of the time of this writing) 
unknown, but it is a positive even function with asymptotics fixed to be $\exp(-\lvert T_r\rvert/ \sqrt{2})$ as $T_r\rightarrow\infty$. Moreover, we have that $\int^{+\infty}_{-\infty}dT \, W(T)=(\textnormal{BPS D}(p-1)\textnormal{-brane tension})$. As noted in \cite{Bergman:2024aly}, integrating $W(T_r)$ from $T_0$ to $+\infty$ yields the fluxbrane charge $\eta$, which depends on the initial value $T_0$ of the tachyon field. As we have already remarked, there is an important difference between holding fixed one particular background value of $T_r$ (in which case we can produce a parameteric family of fluxbrane charges such as those used to engineer topological operators for continuous global symmetries) and the dynamical evolution of $T_r$ in a full string solution.

\section{Transition Between Type I D$8$- and D$7$-Branes } \label{app:TYPEI}

In this appendix we discuss in greater detail some properties of the Type I $\mathrm{D}7$-brane.
As argued in the main body of this paper, this can be viewed as the S-dual of the Het$_{\mathfrak{so}}$ GSO vortex. As mentioned in the main text, the Type I D0-brane experiences a nontrivial Aharonov-Bohm phase around the non-BPS $\mathrm{D}7$-brane \cite{Gukov:1999yn}. Since the D0-brane is in the $\mathbf{S}$ spinor representation of $\mathrm{Spin}(32)/\mathbb{Z}_2^{c}$, this means that there is monodromy of the center $\mathbb{Z}_2\in \mathrm{Spin}(32)/\mathbb{Z}_2^{c}$ along the asymptotic $S^1$ surrounding the D7. In the transverse $\mathbb{R}^2_\perp$ directions, this appears as a $\mathrm{Spin}(32)/\mathbb{Z}_2^{c}$ flux background along $\mathbb{R}^2_\perp$, which is dynamically favored to spread out. This can be confirmed by a scaling of the low-energy action \cite{Schwarz:1999vu}, and agrees with the intuition that a 't Hooft defect does not obey an ``area law'' (more precisely an 8-volume law in this context) unless one is in a Higgs phase. Additional evidence for this instability comes from the fact that there is a localized tachyon on the D7 worldvolume coming from $7-9$ strings.\footnote{Recall that the Type I $\mathrm{D}7$-brane can be built by taking a $\mathrm{D}7-\overline{D7}$ pair in Type IIB string theory whereby the $7-\overline{7}$ tachyon is projected out by $\Omega$.} Indeed the NS sector of the $p-9$ sector for a Type I $\mathrm{D}p$-brane has a ground state energy
\begin{equation}
    M^2=\frac{5-p}{8}.
\end{equation}

In analogy with the $\mathrm{R}7$-brane and its relation to a collapsed IIA/IIB wall wrapped on a contractible circle, our main task in this appendix will be to study a similar relation for the S-dual of the GSO defects of $\mathrm{Het}_{\mathfrak{so}}$, namely the $\mathrm{D}8$- and $\mathrm{D}7$-branes of Type I string theory. More precisely, we will show that the Type I $\mathrm{D}7$-brane is related to a circular configuration of a $\mathrm{D}8$-brane with a monodromy cut for $(-1)^s$, the center of the global form ${\rm Spin}(32)/\mathbb{Z}_2^c$, as in Figure \ref{fig:D7explosion}.

To show this, we first recall the original argument of Sen in \cite{Sen:1998tt} of how the non-BPS $D0$-brane can be built from a $\mathrm{D}1-\overline{\mathrm{D}1}$ pair with a tachyon kink. Recall that the pair has a gauge group $O(1)\times O(1)$ and there is a $1-\overline{1}$ sector tachyon in the representation $(\mathbf{det},\mathbf{det})$ where $\mathbf{det}$ is the nontrivial 1D representation of $O(1)\simeq \mathbb{Z}_2$. We collect the Type I $\mathrm{D}p$ gauge groups in Table \ref{tab:ggrps}. Let us take the spatial worldvolume for the $\mathrm{D}1-\overline{\mathrm{D}1}$ pair to be $S^1$ and let us turn on a $\mathbb{Z}_2$ Wilson line for one of the $O(1)$ factors. Because the tachyon potential $V(T)$ is symmetric under $T\rightarrow -T$, there are two vacuum values $\pm T_0$ for which
\begin{equation}
    V(T_0)=-2g_sT_{D1}
\end{equation}
where $T_{D1}$ is the $\mathrm{D}1$ tension and $g_s$ is the string coupling constant. This means that while the Wilson line precludes the constant solutions $T(\theta)=\pm T_0$ on $S^1$, the minimum-energy configuration will be a tachyon kink due to the boundary condition $T(0)=-T(2\pi R)$. The minimal energy kink solution is topologically protected since $\pi_0$ of the vacuum manifold $\{\pm T_0\}$ is topologically equivalent to $\mathbb{Z}_2$ and is identified with the non-BPS $D0$-brane. 

\begin{table}
  \centering
  $
  \begin{array}{c|c|c|c|c|c|c|c|c|c|c}
D9 & D8 & D7 & D6 & D5 & D4 & D3 & D2 & D1 & D0 & D(-1) \\
\hline O(N) & O(N) & U(N) & Sp(N) & Sp(N) & Sp(N) & U(N) & O(N) & O(N) & O(N) & U(N)
\end{array}
$
  \caption{Gauge groups of Type I D-branes (up to possible $\mathrm{Pin}$-lifts for the $O$ groups and $\mathbb{Z}_2$ quotients for the $\mathrm{Sp}$ groups). Note that $\mathrm{D}p$-branes for $p=6,4,3,2$ are not associated with any $K$-theory charge.}\label{tab:ggrps}
\end{table}

In a similar vein, we can see how the non-BPS $\mathrm{D}7$-brane can be obtained from \textit{two}\footnote{We will connect with the configuration of Figure \ref{fig:D7explosion} involving a single $\mathrm{D}8$ at the end of the section.} $\mathrm{D}8$'s with a Wilson line. The reason we need two $\mathrm{D}8$'s follows from the fact that the gauge group of the $\mathrm{D}7$ is $U(1)$ while 
that of a single $\mathrm{D}8$ is $O(1)$, so we cannot Higgs it to the former. Further, two $\mathrm{D}8$'s have a gauge group of $O(2)$ and an $8-8$ tachyon in the $\mathbf{det}$ representation which we denote $T_{\mathbf{det}}$. In terms of 2x2 matrices, this tachyon field can be written as
\begin{equation}\label{eq:tachyonpmatrix}
    \begin{pmatrix}
        0 & T_{\mathbf{det}} \\
        -T_{\mathbf{det}} & 0.
    \end{pmatrix}
\end{equation}
A vev for $T_{\mathbf{det}}$ will Higgs the $O(2)$ gauge group to the commutator subgroup $U(1)$. Without any background Wilson line on this $\mathrm{D}8$-stack, the entire system would just annihilate, but the presence of an $O(2)$ Wilson line with monodromy 
\begin{equation}\label{eq:O(1)subsetO(2)Wilsonline}
   \begin{pmatrix}
       -1 & 0\\
       0 & 1
   \end{pmatrix}
\end{equation}
around an $S^1$ will produce a topologically stable kink configuration for $T_{\mathbf{det}}$ similar to what we saw in the previous paragraph. This produces a 7-brane in Type I with a localized $U(1)$ gauge field for which F1 strings can end because they can end on the D$8$ constituents. The only known candidate for such a brane would be the Type I D$7$-brane. To confirm this explicitly, one can start with two non-BPS D$8$ branes in IIB, considering a tachyon kink configuration which creates a D$7$-$\overline{\mathrm{D}7}$ pair and seeing if this matches the $T_{\mathbf{det}}$ upon projecting to Type I.\footnote{This matching would show to claim because it is known that a single D$8$-brane in Type IIB becomes a single D$8$-brane after the $\Omega$ projection, while a D$7$-$\overline{\mathrm{D}7}$ pair becomes a single D$7$-brane after the projection.} Such a tachyon kink in the IIB D$8$ brane system is valued in the adjoint of $U(2)$, which is the gauge group of the D$8$ brane system, and a kink 
\begin{equation}
    T=\begin{pmatrix}
    T_0 \tanh{(x_{\perp}/B)} & 0 \\
    0 & -T_0 \tanh{(x_{\perp}/B)}
    \end{pmatrix}
\end{equation}
for some width $B$ produces a D$7$-$\overline{\mathrm{D}7}$ pair by a simple consideration of the diagonal subgroup $U(1)\times U(1)\subset U(2)$ which lives on separated non-BPS D$8$-branes. By a $U(2)$ gauge transformation, we can rotate the tachyon profile to
\begin{equation}\label{eq:u2tachyon}
    T=T_0 \begin{pmatrix}
        0 & -iT_0 \tanh{(x_{\perp}/B)} \\
        iT_0 \tanh{(x_{\perp}/B)} & 0
    \end{pmatrix}.
\end{equation}
Upon projecting to Type I, the D$8$ gauge group is projected as $U(2)\rightarrow O(2)$ and the adjoint of $U(2)$ becomes the adjoint of $O(2)$ which happens to be the $\mathbf{det}$ representation. We indeed see that \eqref{eq:u2tachyon} matches \eqref{eq:tachyonpmatrix} up to an overall factor of $i$, showing that the claim that we have produced the Type I D$7$-brane.\footnote{We would also claim that, by Bott periodicity, a similar correspondence can be established between two (Euclidean) Type I $D0$-branes on an $S^1$ with a Wilson line \eqref{eq:O(1)subsetO(2)Wilsonline} inducing a tachyon kink which is equivalent to a $D(-1)$-brane.} This in particular implies that the $O(2)$ Wilson line of the D$8$ system can be an endpoint for a bulk monodromy cut for $(-1)^s$ as it sources the D$7$-charge. 

One can then obtain a configuration similar to right side of Figure \ref{fig:D7explosion} but with two $\mathrm{D}8$-branes in the circular configuration instead of one. The fact that the above monodromy can be presented as an $O(1)$ monodromy means that we can simply annihilate one of the $\mathrm{D}8$'s in the interior of the circle, yielding the configuration of Figure \ref{fig:D7explosion}. 

Finally, we close this appendix with mentioning how a single $\mathrm{D}8$-brane is expected to decay. While there is no tachyon in the $8-8$ sector for a single such brane, there is a tachyon in the $8-9$ sector which is in the representation $(\mathbf{det},\mathbf{32})$ of $O(1)\times O(32)$ \cite{Schwarz:1999vu}. Here $O(1)$ is the gauge group localized on the $\mathrm{D}8$, while $O(32)$ is the flavor symmetry of the $\mathrm{D}8$ which is an imprint of the bulk $\mathfrak{so}(32)$ gauge theory. Let us denote this tachyon\footnote{One could ask what role, if any, such a tachyon played in the above configuration of two $\mathrm{D}8$-branes on a circle with an $O(2)$ Wilson line. In this setup, the $8-9$ sector tachyon is in the representation ($\mathbf{2}$,$\mathbf{32}$) of $O(2)\times O(32)$. While the vector $\mathbf{2}\rightarrow -\mathbf{2}$ under the Wilson line, the $\mathrm{D7}$ sourced by this setup induces a bulk monodromy in $\mathbb{Z}^s_2$ under which $\mathbf{32}$ is odd and therefore the $8-9$ sector tachyon is invariant under the $O(2)$ Wilson line. Indeed these tachyons just become the known $7-9$ tachyons on the $\mathrm{D}7$-brane worldvolume.}, $T_i$, $i=1,...,32$. Then its potential $W(T_i)$ will be invariant under the gauge and flavor symmetry transformations:
\begin{equation}
    T_i\rightarrow -T_i \;  \; \mathrm{or} \; \;  T_i\rightarrow R^{SO(32)}_{ij}T_j \; \implies W\rightarrow +W
\end{equation}
This means that inside the scalar field space $\mathbb{R}^{32}/\mathbb{Z}_2$, there is a vacuum submanifold $\mathbb{RP}^{31}$ where $W$ is at its minimum value, $-g_sT_{\mathrm{D8}}$. The initial state of a $\mathrm{D}8$ is characterized by the local maximum $W(T=0)=0$ and dynamically different regions, labeled by $R$, of the $\mathrm{D}8$ will roll down to various vacuum values $p_R\in \mathbb{RP}^{31}$ and the entire $\mathrm{D}8$ will dilute away.

\newpage

\bibliographystyle{utphys}
\bibliography{ABWall}

\end{document}